\documentclass[traditabstract,a4paper]{aa}
\usepackage{txfonts} 

\usepackage{color}

\usepackage{natbib}
\bibpunct{(}{)}{;}{a}{}{,} 

\usepackage{graphicx}
\usepackage{fixltx2e}

\usepackage{longtable}
\usepackage{multirow}
\usepackage{overpic}
\usepackage{array}

\definecolor{RoyalPurple}{cmyk}{0.75,0.9,0,0.1}
\definecolor{MyBlue}{rgb}{0.0025,0.1125,0.2}
\usepackage[breaklinks, colorlinks, citecolor=blue, linkcolor=MyBlue, urlcolor=RoyalPurple, 
            colorlinks=true, linkcolor=blue, debug, baseurl=' ']{hyperref}

\graphicspath{{Figures/}}

\begin{document}

\title{LEKID sensitivity for space applications between 80 and 600~GHz}

\author{
A.~Catalano \inst{1},
A.~Bideaud \inst{2},
O.~Bourrion \inst{1},
M.~Calvo \inst{2},
A.~Fasano \inst{2},
J.~Goupy \inst{2},
F.~Levy-Bertrand \inst{2},
J.F.~Mac\'{\i}as-P\'erez \inst{1},
N.~Ponthieu \inst{3},
Q.Y. Tang \inst{2,4}
\and
A.~Monfardini \inst{2}
}

\offprints{A. Catalano - catalano@lpsc.in2p3.fr} 

\institute{Univ. Grenoble Alpes, CNRS, LPSC/IN2P3, 38000 Grenoble, France
\and Univ. Grenoble Alpes, CNRS, Grenoble INP, Institut N\'eel, 38000, Grenoble France
\and Univ. Grenoble Alpes, CNRS, IPAG, 38000 Grenoble, France
\and
Kavli Institute for Cosmological Physics, University of Chicago, 60637, Chicago, IL, USA
}  
 
 \abstract
 {We report the design, fabrication and testing of Lumped Element Kinetic Inductance Detectors (LEKID) showing performance in line with the requirements of the next generation space telescopes operating in the spectral range from 80 to 600~GHz. This range is of particular interest for Cosmic Microwave Background (CMB) studies. For this purpose we have designed and fabricated 100-pixel arrays covering five distinct bands. These wafers have been measured via multiplexing, where a full array is read out using a single pair of lines. We adopted a custom cold black-body installed in front of the detectors and regulated at temperatures between 1~K and 20~K. We will describe in the present paper the main design considerations, the fabrication processes, the testing and the data analysis.}
 
\authorrunning{A. Catalano et al.}

\maketitle

\section{Introduction} \label{sec1}
Instruments using Kinetic Inductance Detectors (KIDs) have been demonstrated to reach state-of-the-art performance in case of ground-based observations. The kilo-pixel dual-band NIKA2 camera was the first KID-based instrument operating at millimetre wavelengths. It showed that LEKID detectors are limited by photon noise when they operate under an optical load typical of ground-based observations \citep{nika2_1,nika2_2}. The last space-borne instrument devoted to Cosmic Microwave Background (CMB) observations, Planck, has used 52 high impedance spider-web bolometers  in the high frequency instrument (HFI), and already showed their sensitivities were limited by CMB photon noise \citep{planck_1,planck_2,planck_3,planck_4,planck_5}. Despite the fact that HFI bolometers had the best sensitivity possible per pixel, after five full sky surveys, the CMB B-mode \citep{1997PhRvD..56..596H} has been only marginal detected by Planck. For this reason, new proposed space missions \citep{2016AAS...22830108L} and the upcoming CMB Stage IV ground experiments \citep{2020AAS...23523504T,2019BAAS...51g.209C} aim improve the overall Noise Equivalent Power (NEP) of new instruments by increasing the focal plane coverage, using thousands of photon noise-limited contiguous pixels. 

In this paper, we present our recent developments concerning the design, fabrication and testing of complete arrays of LEKID operating between 80 and 600~GHz and under an optical loading typical of the space environment. This particular range of frequencies has been chosen because it includes the peak of the CMB emission (around 150~GHz), and it extends sufficiently to ensure an efficient measurement and subtraction of the foregrounds. In order to focus our efforts, the arrays have been optimised for five different bands: 80-120~GHz (centred on 100~GHz), 120-180~GHz (centred on 150~GHz), 180-270~GHz (centred on 240~GHz), 320-400~GHz (centred on 360~GHz) and 450-650~GHz (centred on 550~GHz). 
Each array consists of around a hundred pixels and is multiplexed so that a single readout line can read out all the pixels on an array.
In Sec.~\ref{sec2} we present the requirements and the experimental setup. In Sec.~\ref{sec3} we illustrate the main design guidelines and the fabrication processes. In Sec.~\ref{sec4} we describe the laboratory tests that permitted extensive device characterisation, in particular in terms of the sensitivity, spectral and dynamical performance.

\begin{figure}
\begin{center}
\includegraphics[width=6cm, angle=-90]{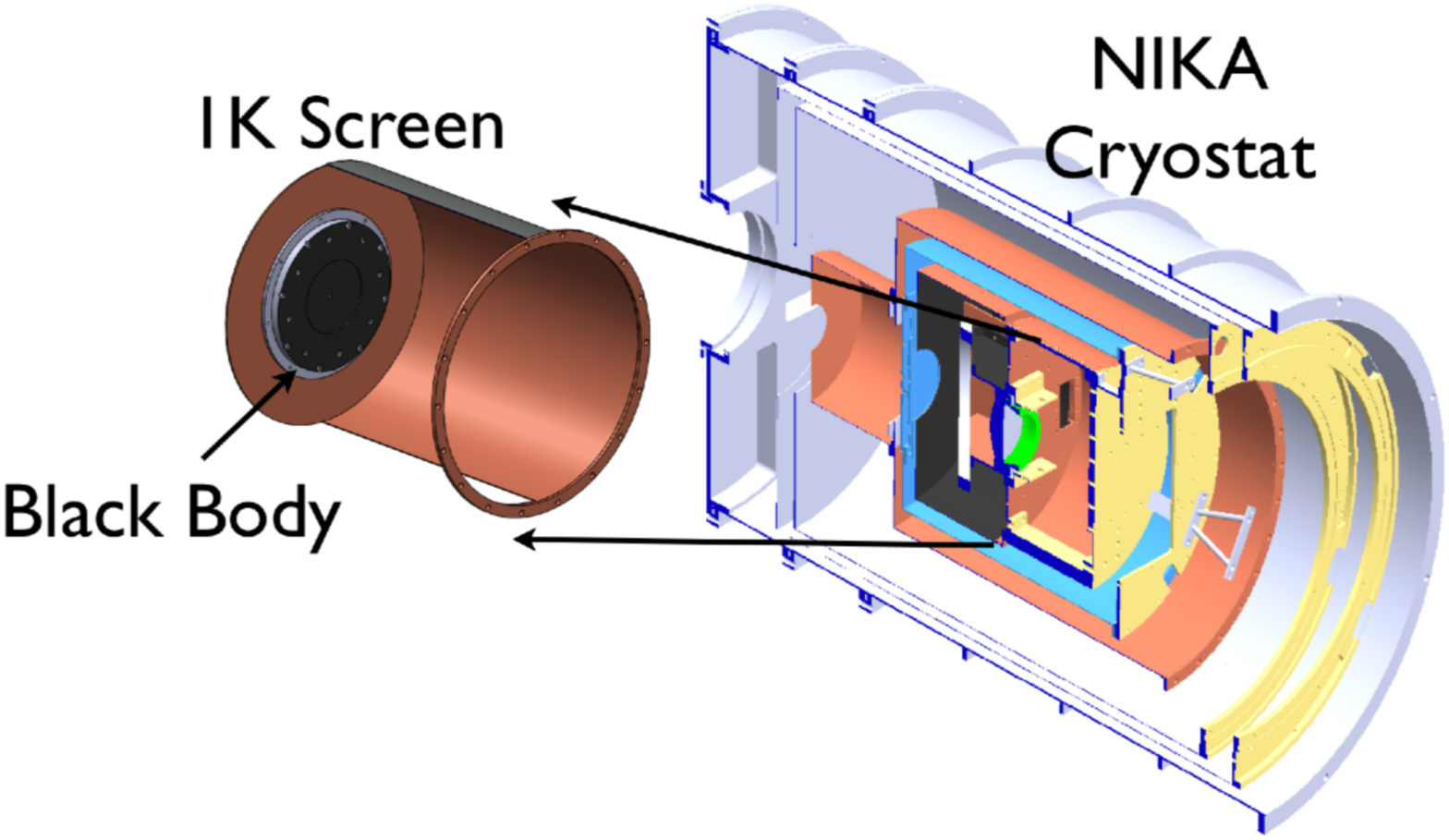}
\includegraphics[width=6cm,angle=-90]{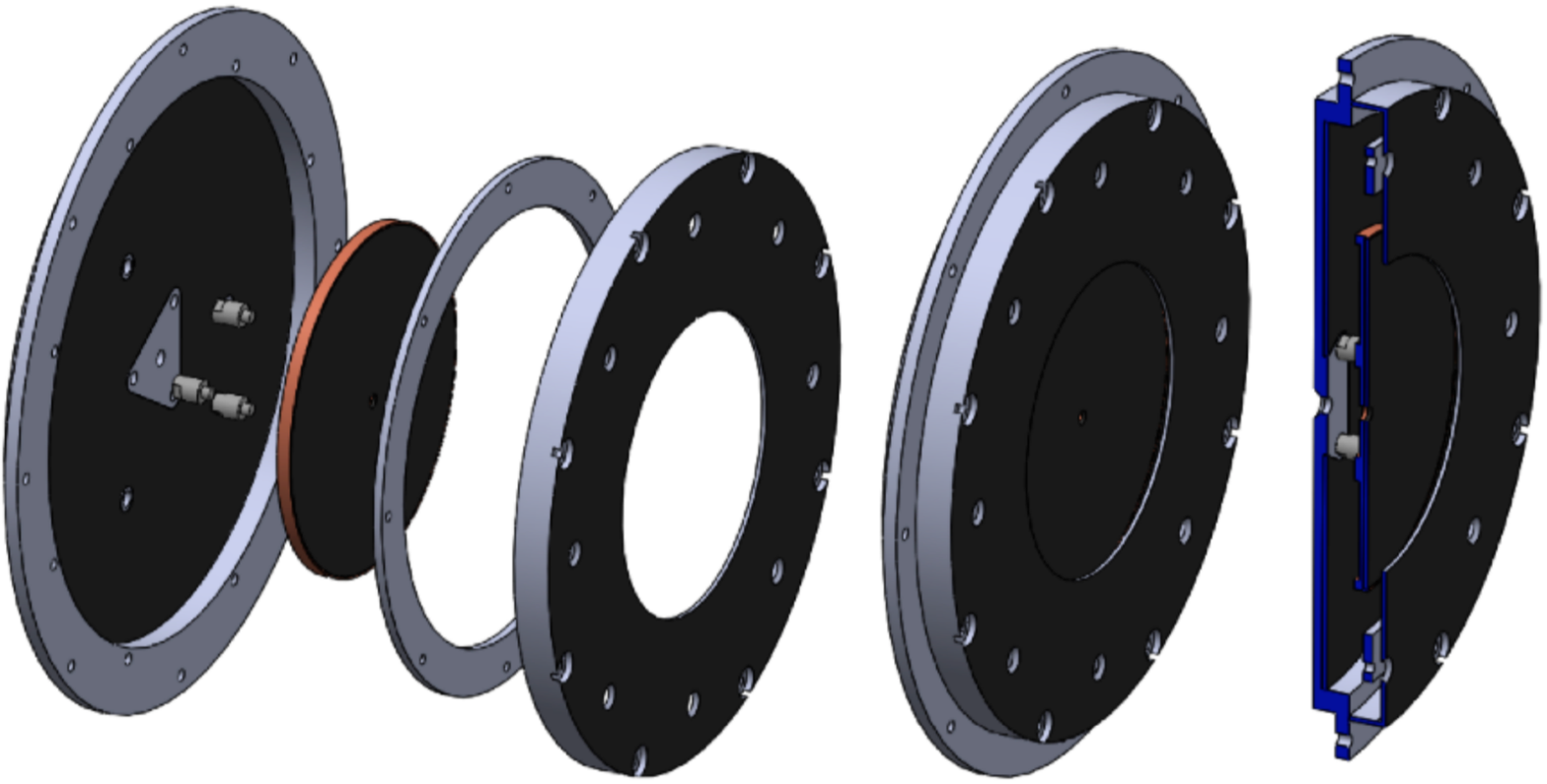}
\end{center}
  \caption{Top: 3D drawing of the dilution cryostat used for the tests. The black-body is mounted at the end of the still screen, at a base temperature of around 1~K. Bottom: Closeup view of the cryogenic back-body.}
\label{fig1}
\end{figure}

\section{Requirements and experimental setup}\label{sec2}

When operated above the atmosphere, an instrument covering the band 80-600~GHz is loaded mostly by the 2.72~K CMB and the thermal radiation of the interstellar dust. In addition, above roughly 300~GHz, the instrument itself becomes an important source of optical background due to the thermal emission of the telescope \citep{planck_6}. For typical configurations (e.g. Planck, LiteBIRD \citep{2019BAAS...51g.286L}, or other works \citep{2017A&A...601A..89B}, \citep{2018JCAP...04..015D}), the optical loading on a pixel is in the range of 0.1-1~pW, depending on the observed bandwidth and other factors such as the instrumental throughput. The needed noise equivalent power (NEP) is at least one order of magnitude lower compared to the ground-based case. For this work, we define our goal NEP per pixel as:

\begin{equation}\label{eq1}
NEP_{goal}  \leq \sqrt{2} \cdot NEP_{phot}
\end{equation}

where $NEP_{phot}$ represents the source
of noise related to the photon noise\footnote{The photon noise comes from the intrinsic fluctuations of the incident radiation that are due to the Bose-Einstein distribution of the photon emission. In the best-case scenario, the contribution of photon noise is the limit of sensitivity of the instrument}. $\sqrt{2}$ means that we require that all the other sources of instrumental noise as random quasiparticle recombination, TLS noise \citep{2016ApPhL.108h3504F} and pre-ampli noise to not exceed the photon noise level. When this condition is satisfied, we refer to the devices as photon-noise dominated detectors.

In order to reproduce the particular conditions of optical power for each band, we used a dilution cryostat operating at 100~mK and equipped with a cryogenic black-body installed on the 1~K (still) stage. The black-body can be regulated, via a standard PID procedure, between 1~K and 20~K. This has no impact on the thermal stability of the coldest stage and the detectors. The black-body (see Fig.~\ref{fig1}) is composed of: a) an aluminium support covered by black stycast glue, b) three Vespel legs, c) a copper disk covered by carbon-doped black stycast mixed with SiC grains, and d) a metallic support to hold the band-defining filters. The copper disk, with its highly emissive coating, represents the actual black body. The performance of the coating has been compared with the commercial foam \emph{Eccosorb AN-72}, rated for better than -17~dB reflectivity at $\geq$20~GHz. The measurement, performed at 150~GHz, show comparable emissivity for both cases.
The black body geometry has been studied in order to minimise the thermal capacitance and ensure a good isotherm behaviour of the emissive part. The Vespel legs have a purely mechanical role, and presents a negligible thermal conductance. A calibrated, and easily replaceable, thermal link is thus added to adjust the time constant. In order to avoid multiple reflections which could lead to increase the final throughput, we fully coated the cryogenic screens between the black body and the detectors. In addition, the 1-Kelvin (still) screen is optically closed with respect to the warmer screen in order to avoid stray lights.
\\
The five spectral bands are defined by low-pass, high-pass, or band-pass metal mesh filters. For the detectors covering the 3~mm and 2~mm bands the low-frequency cutoff is provided by the superconductor film itself (see Sec.~\ref{subsec4-1}). 

The optical loading per pixel is equal to:

$$
W_{pixel} = \tau \int_{0}^{\infty} A \Omega \chi (\nu) BB(T,\nu) d\nu
$$

Where:

- $A \cdot \Omega$ is the throughput of the system. A is the area of the 100~mK cold stop, through which the black-body is observed. $\Omega$ is the solid angle between the cold stop and the pixel.


- $\tau$ is the overall transmission, mostly the transmission of the optical filters used to select the spectral bands. This parameter has been characterised by the Cardiff team who provided the filters and its value is between 90 and 95~\%. The typical out-of-band rejection, confirmed also by in-house measurements, is better than 1~\% per filter. The effective high-frequency rejection is further enhanced by the fact that these pixels, mostly optimised for millimeter-waves, are by construction very poor THz-absorbers.

- $\chi(\nu)$ is the measured spectral band, normalised to 1 at the peak. By this normalisation we assume that the absorption is 100~\% at the peak. This assumption has been demonstrated by reflection measurements using a network analyser, an rf-setup to provide the frequency band, a feed horn, a corrugated lens and the array (with its back-short) \citep{Roesch}. 

- BB(T,$\nu$) is the brightness of the black-body at the working temperature. We assume an ideal emissivity across the whole band.

\begin{figure}
\begin{center}
\includegraphics[trim={2cm 13cm 0cm 2cm}, clip=true,scale=0.5, angle=0]{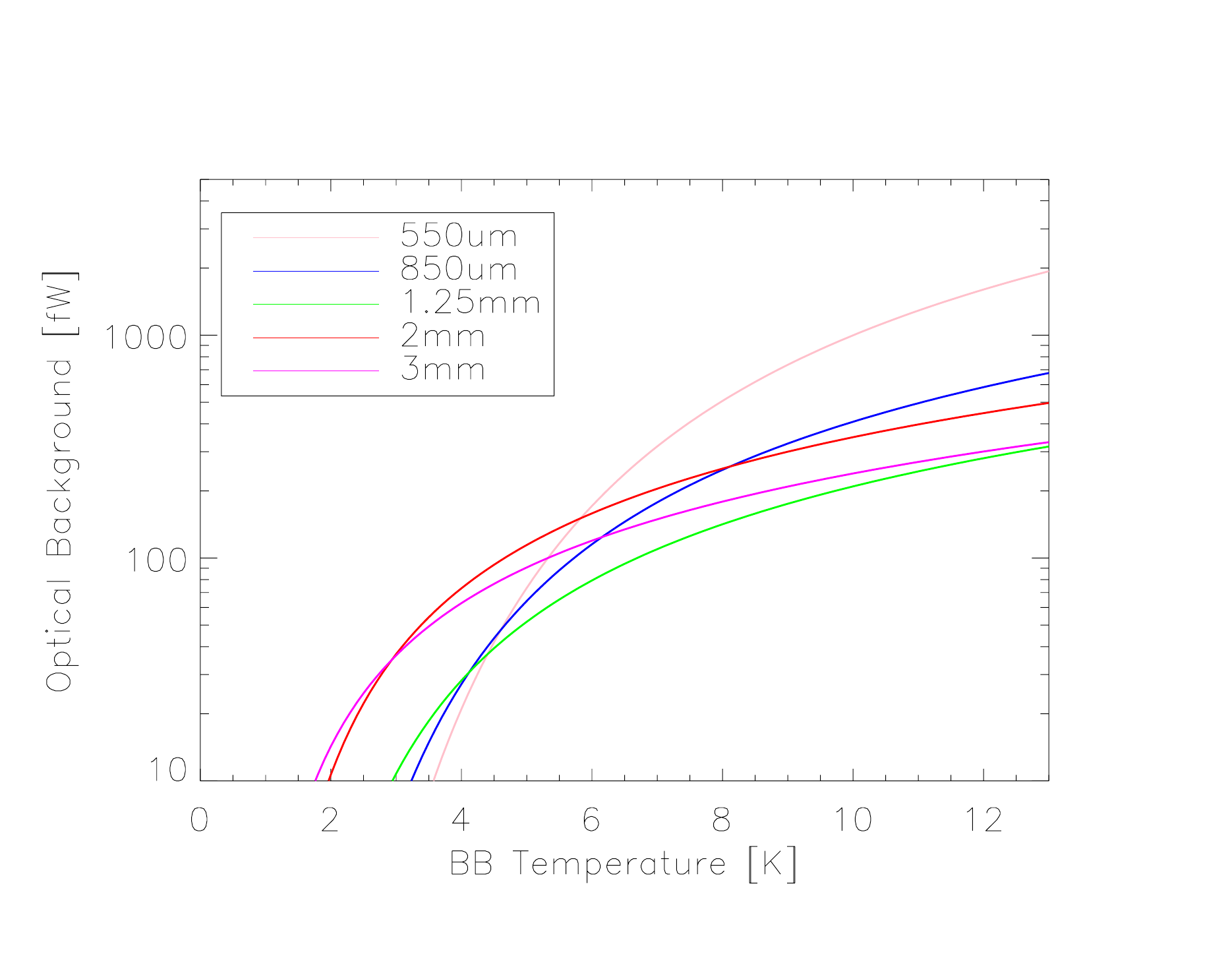}
 \caption{Estimation of the optical power per pixel as a function of the temperature of the cold black-body.  
\label{fig2}}
\end{center}
\end{figure}

In Fig.~\ref{fig2}, we present the results of an optical simulation performed for this setup. In particular, we show the optical power per pixel integrated in the specific spectral band. 

\section{Design and fabrication of the LEKID arrays}\label{sec3}

For this study, we have specifically designed and fabricated arrays optimised for low background, i.e. 0.1-1~pW per pixel, and for five distinct spectral bands centred on wavelengths of 3, 2, 1.25, 0.85, and 0.55~mm. The final designs are the result of extensive electro-magnetic (transmission-line model, planar and full 3-D) calculations and simulations. The detectors that have been used are back-illuminated LEKID, i.e. LC resonators made by a long meandered inductor terminated at both ends by an interdigitated capacitor. Most of the arrays are based on thin aluminum films (t$_{FILM}$ = 15-18~nm), with a critical temperature T$_c \approx$~1400~mK, corresponding to a lower frequency cutoff of about 110~GHz. On the other hand, the array operating at 3 mm employs a titanium-aluminum bilayer film (t$_{FILM}$ = 35~nm, with t$_{Ti}$ = 10~nm and t$_{Al}$ = 25~nm) with T$_c$ = 900 $\pm$ 25~mK. For a standard Mattis-Bardeen superconductor, this results in a theoretical cutoff around 70~GHz \citep{catalano_1}. The LEKID planar design is based mostly on results from adaptations and optimisations starting from the pixels elaborated for the NIKA instrument \citep{nika1}. Among the most important resonators parameters to be optimised out of planar simulations is the coupling quality factor of the resonators which has been adjusted, for the present study, to be in the range of Q$_c \approx 5\cdot10^4-10^5$. 

\begin{figure}
\begin{center}
\includegraphics[width=9cm, angle=0]{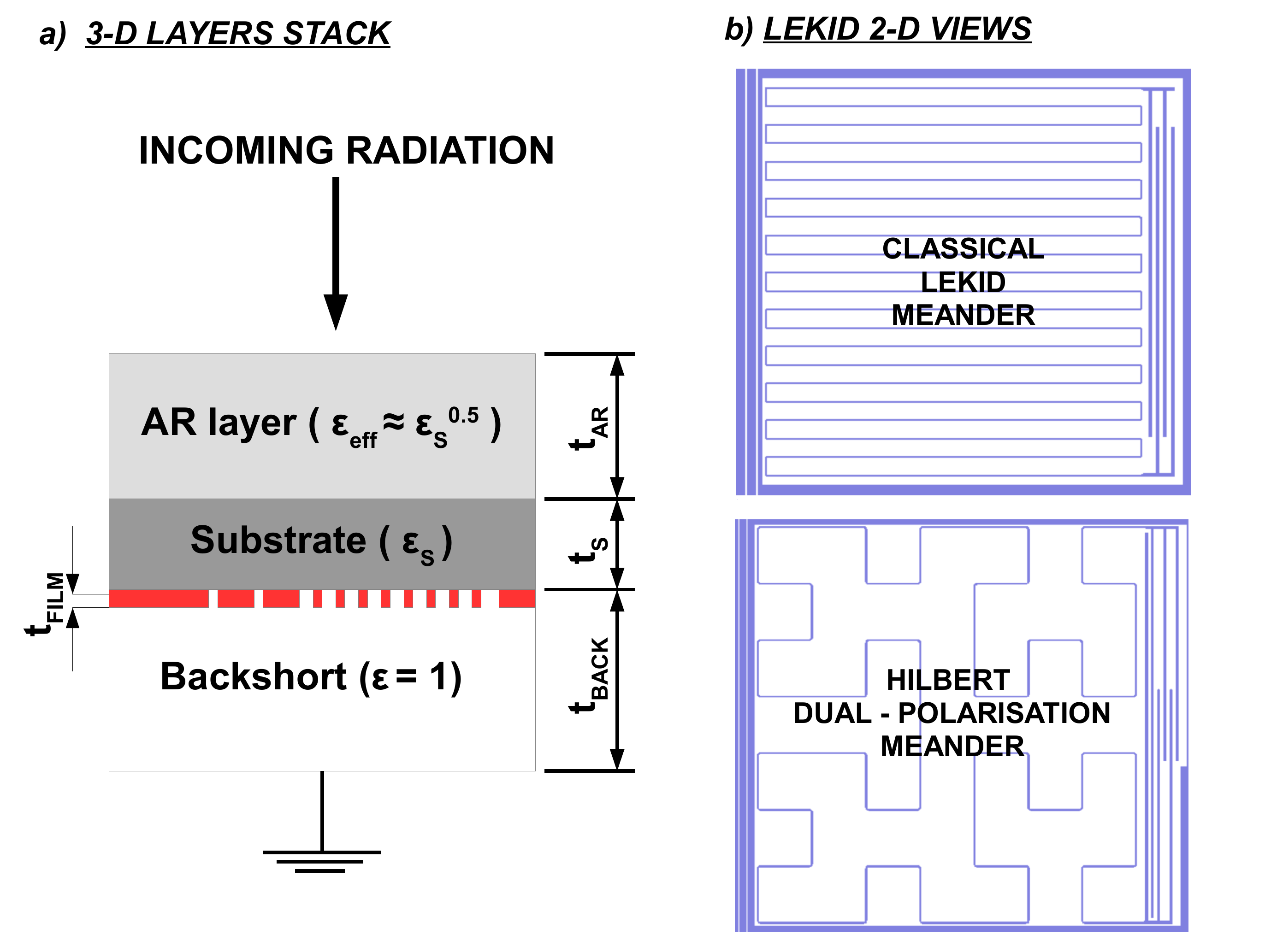} \\
\includegraphics[width=6cm, angle=0]{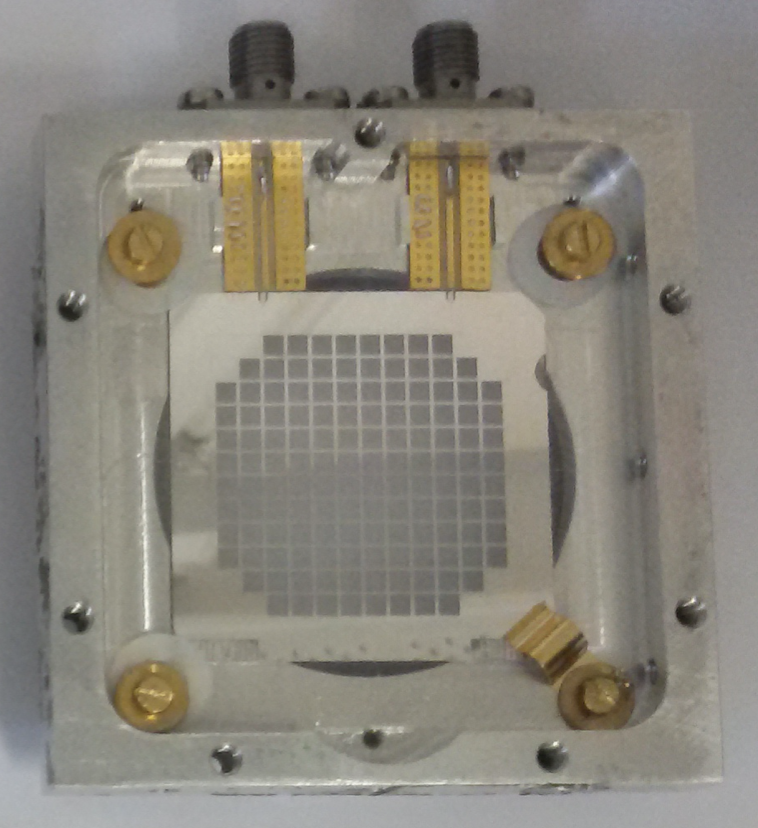}
\end{center}
  \caption{a) LEKID 3-D stack. In red, the superconducting film (thickness not-to-scale). The micromachined AR has been made using only silicon substrates. The backshort lid is superconducting to avoid affecting the quality factor of the resonators; b) the two kinds of planar LEKID designs adopted: the classical (CL) meander or the Hilbert (H) dual-polarisation structure. Bottom panel: picture of the final prototypes at 3~mm.}
\label{fig3}
\end{figure}

\begin{figure}[h]
\begin{center}
\includegraphics[width=9cm, angle=0]{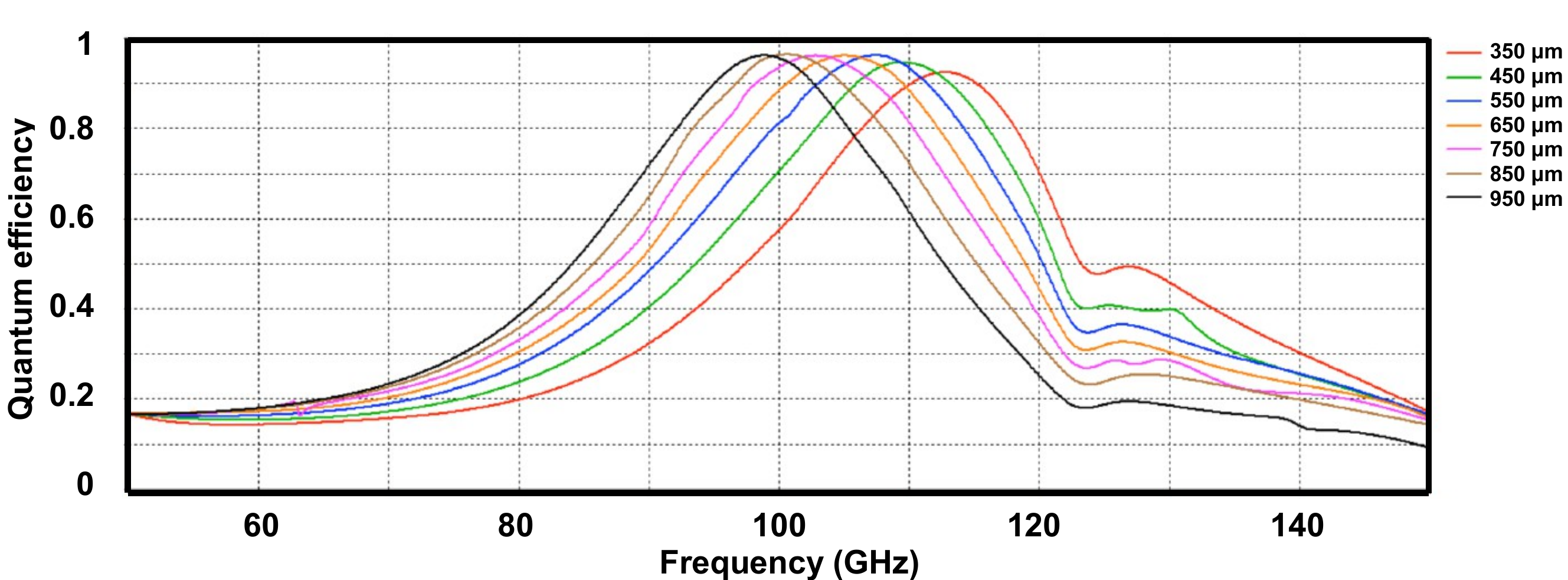}
\end{center}
\caption{Transmission-line model calculation to optimise, in terms of the pixel quantum efficiency, the backshort distance t$_{BACK}$ for the 3~mm array. The optimum value is established at t$_{BACK}$~$\approx$~750~$\mu$m.}
\label{fig4}
\end{figure}

In the Fig.~\ref{fig3}, we present the three dimensional layout and examples of the LEKID planar views. As far as the optical coupling efficiency is concerned (left part of Fig.~\ref{fig3}), the key variables to be calculated include the substrate (high-resistivity HR silicon or sapphire) thickness t$_{S}$, the thickness and effective dielectric constant of the anti-reflecting (AR) layer t$_{AR}$, $\epsilon _{AR}$ (only for silicon substrates), the thickness of the vacuum backshort t$_{BACK}$ and the effective impedance of the superconducting (Al or Ti-Al) film as seen by the incoming wave. This last parameter is strongly related to the superconducting material(s) properties, and the thickness of the film t$_{FILM}$. Fig.~\ref{fig4} shows an example of a transmission-model estimation of the optical efficiency of the 3~mm array when varying t$_{BACK}$. 

The 2-D planar structures printed in the superconducting film (right part of Fig.~\ref{fig3}) are designed to ensure a good optical coupling and satisfy the circuit requirements in terms, for example, of quality factor and resonance frequency f$_0$. For the present study, we have adopted the resonance frequency range of f$_0$ = 1-3~GHz. The main geometrical characteristics of the arrays are presented in Tab~\ref{tab1}. 

\begin{table}[b!]
\scriptsize
\begin{center}
\begin{tabular}{cccccc}
\hline
\hline
Wavelength & 3~mm  & 2~mm & 1.25~mm & 850~$\mu$m & 550~$\mu$m \\
Frequency & 100~GHz  & 150~GHz & 240~GHz & 360~GHz & 550~GHz \\
\hline \hline
Film material & Ti-Al    & Al & Al & Al & Al \\
Film thick. [$n m$] & 10-25 & 18 & 15 & 15 & 15 \\
Critical temp. [K] & 900 & $\approx$ 1.3 &1.45&1.45& 1.45\\
Substrate material & Al$_{2}$O$_{3}$ & Al$_{2}$O$_{3}$ &Si&Si&Si\\
Substrate thick. [$\mu$m] & 430 & 330 & 280 & 265 & 170 \\
AR layer thick. [$\mu$m] & 0 & 0 & 110 & 110 & 80 \\
Backshort thick. [$\mu$m] & 750 & 670 & 530 & 350 & 400 \\
LEKID meander type &   CL & H & H & CL & CL \\
Pixel size [mm] &  2.3 & 2.3 & 1 & 1 & 1 \\
Design pixels/array [\#] & 132 &  132 & 132 & 128 & 128 \\
\hline \hline
\end{tabular}
\end{center}
\caption{Main characteristics of the arrays.}
\label{tab1}
\end{table}

The fabrication process is similar for all arrays. The substrate (either HR~Si or C-cut Al$_{2}$O$_{3}$) is prepared by applying an HF vapour cleaning. The superconducting film is deposited by e-beam evaporation and under a residual chamber pressure of 5$\cdot$10$^{-8}$~mbars. In the case of the 3~mm array, the titanium (first) and aluminum (second) layers are deposited in sequence without breaking the vacuum. The deposition rates are within the range 0.1-0.5~nm/sec. This procedure guarantees a clean interface, and the protection of the (underlying) Titanium layer when the chip is exposed to the atmosphere.

The UV photo-lithography step is achieved using a positive resist, and is followed by wet etching using a standard aluminum etching solution based on a diluted H$_{3}$PO$_{4}$/HNO$_{3}$ mixture. A dip in 0.1~\% HF, to etch the underlying titanium film, is added for the 3~mm array case. The arrays made on Silicon substrates undergo the AR-layer step by dicing, on the back of the Si wafer, a regular cross-pattern of grooves \citep{dicing2016}.


The diced and eventually AR-coated detectors are packaged in custom holders and bonded using superconducting aluminum wires (dia. 17~$\mu m$). The detector is connected via SMA connectors to high-quality coaxial cables and a cold low-noise amplifier to be readout with the external multiplexing circuit \citep{nikel2012}.

\section{Detectors characterisation}\label{sec4}

The arrays described in Sec.~\ref{sec3} were first tested with the cryostat closed and the detectors illuminated by the cold black-body (see Sec.~\ref{sec2}). The results of these measurements are shown in the Sec.~\ref{subsec4-1} and \ref{subsec4-3}. In order to correctly establish a NEP (Noise Equivalent Power) calculation, we have spectrally characterised the very same arrays. For this measurement, the full cryostat optics is installed, and the camera is interfaced with the Fourier Transform interferometer. These experiments are described in more detail in Sec.~\ref{subsec4-2}.
In addition, the 3~mm array was tested with our fast, i.e. 1~MHz sampling, homodyne electronics to determine the LEKID response time constant (see Sec.~\ref{subsec4-4}). For space instruments which are exposed to a considerable flux of primary cosmic rays, the response time is a crucial parameter \citep{Masi_Olimpo2019,catalano_2}.

\subsection{Resonators electrical properties characterisation}\label{subsec4-1}

The 132 LEKID of the 3~mm (Ti-Al, t$_{FILM}$~=~10-25~nm) and the 2~mm (Al, t$_{FILM}$~=~18~nm) arrays (pixels size 2.3~mm) resonate between 1.4 and 1.9~GHz. The arrays have average internal quality factors of 4.8x10$^4$ and 6.2x10$^4$, respectively, under typical CMB loading (see first raw in Tab.\ref{tab2}). The critical temperature of the Ti-Al film has been roughly estimated, in this case, by monitoring the readout line transmission, at around 900~mK. The transition is crossed multiple times in both ways (warming-up and cooling-down) in order to exclude potential thermal drifts. This value is in agreement with previous measurements on similar films \citep{catalano_1}.

The normal-state sheet resistance of the Ti-Al bilayer has been independently measured at T=1~K to be $R_N^{Ti-Al} = 0.5~\Omega$/square.

The 132 LEKID of the 1~mm (Al, $t_{FILM}=15nm$) array (pixels size 1~mm) resonates between 1.7 and 2.1~GHz. These detectors were measured to have an average internal quality factor of 5.0x10$^4$. The superconducting transition temperature of the 15~nm film is 1450$\pm$50~mK, while the normal-state sheet resistance is R$_N^{Al} = 2~\Omega$/square.

The normal-state sheet resistance (measured) is related to the critical temperature (measured) and the surface inductance through the relation \cite{leduc}, $L_s$~=~$\frac{\hbar R_s}{\pi \Delta,}$, where $\Delta$ is the superconducting gap equal to $\Delta = 1.764 \cdot$~k$_B \cdot$~T$_c$ for BCS ideal superconductors. The surface inductance has been estimated from electro-magnetic simulation, in order to explain the observed position of the fundamental resonance frequency. In the case of the Ti-Al film, we found L$_s \approx$~1~pH/square. The three electrical values ($\Delta$, L$_s$ and R$_s$), measured or estimated with independent methods, satisfy the cited relation within 10~\%.

\subsection{Spectral characterisation}\label{subsec4-2}

\begin{figure}[h]
\begin{center}
\includegraphics[width=8.5cm, angle=0]{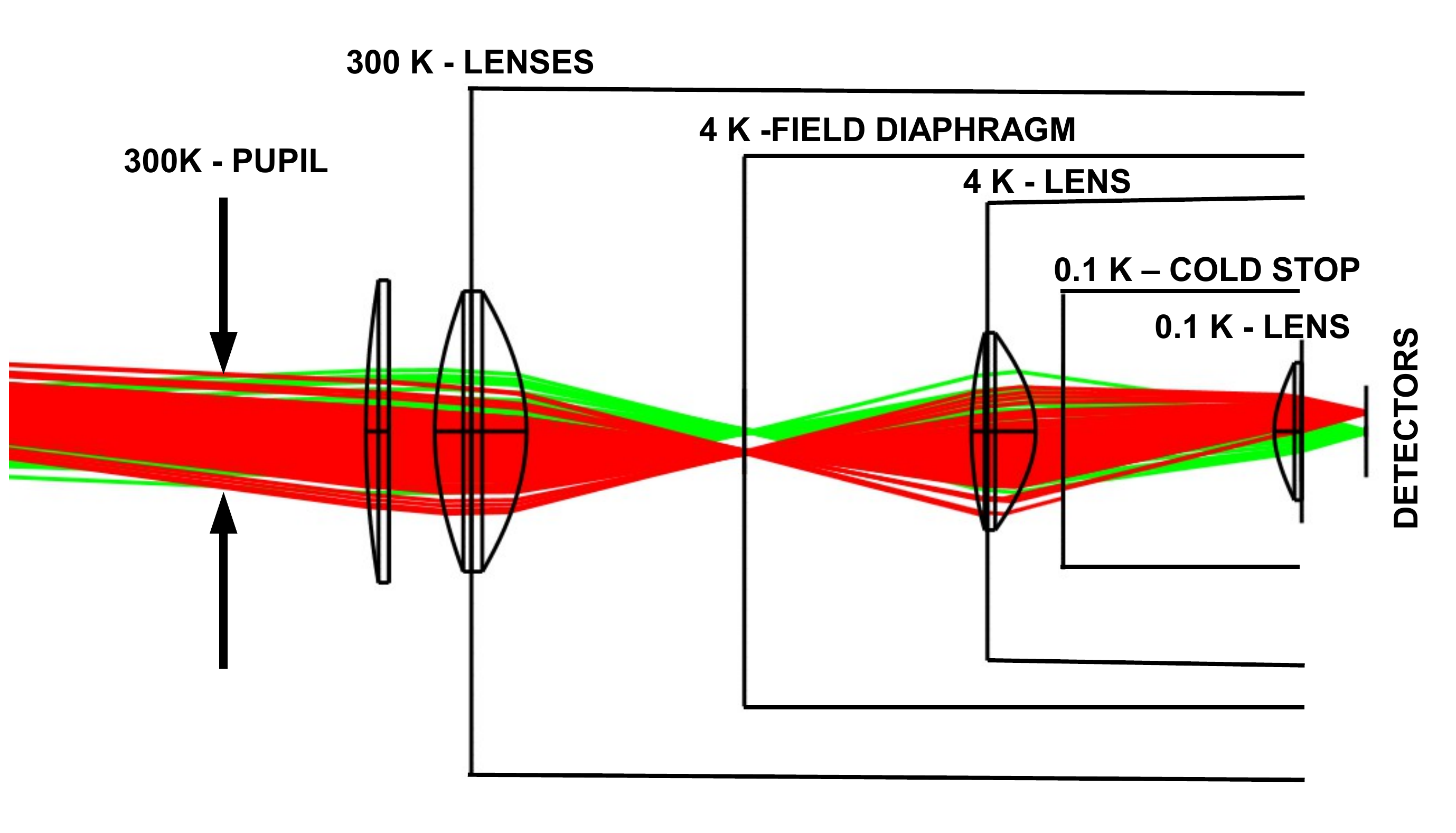}
\end{center}
\caption{Scheme of the optics adopted for the spectral characterisation. The lenses are made out of HDPE (High Density Polyethylene). The low-pass and thermal filters positions are not shown. The interferometer input window is located at the 300~K pupil position.}
\label{fig5}
\end{figure}

The spectral response has been measured using a custom built Martin-Puplett Interferometer (MpI) interfaced to the camera (cryostat) containing the detectors. The optics configuration for this measurement is shown in Fig.~\ref{fig5}. Infrared-blocking filters are installed on the 300~K, 150~K and 50~K cryogenics stages. Metallic multi-mesh low-pass filters are placed at 50~K, 4~K and 1~K. The band-defining filters, again metallic multi-meshes, are installed at base temperature, i.e. around 100~mK, in front of the array.

In order to maximise the signal to noise ratio of the interferogram, we modulate the input signal between two distinct Rayleigh-Jeans spectrum sources held at different temperatures. The modulation is achieved by a rotating wire-grid polariser which also produces the fully polarised input signal. The signal is then extracted with a standard lock-in detection.  

\begin{figure*}
\begin{center}
\includegraphics[trim={3.5cm 14cm 0cm 4cm}, clip=true,scale=0.52]{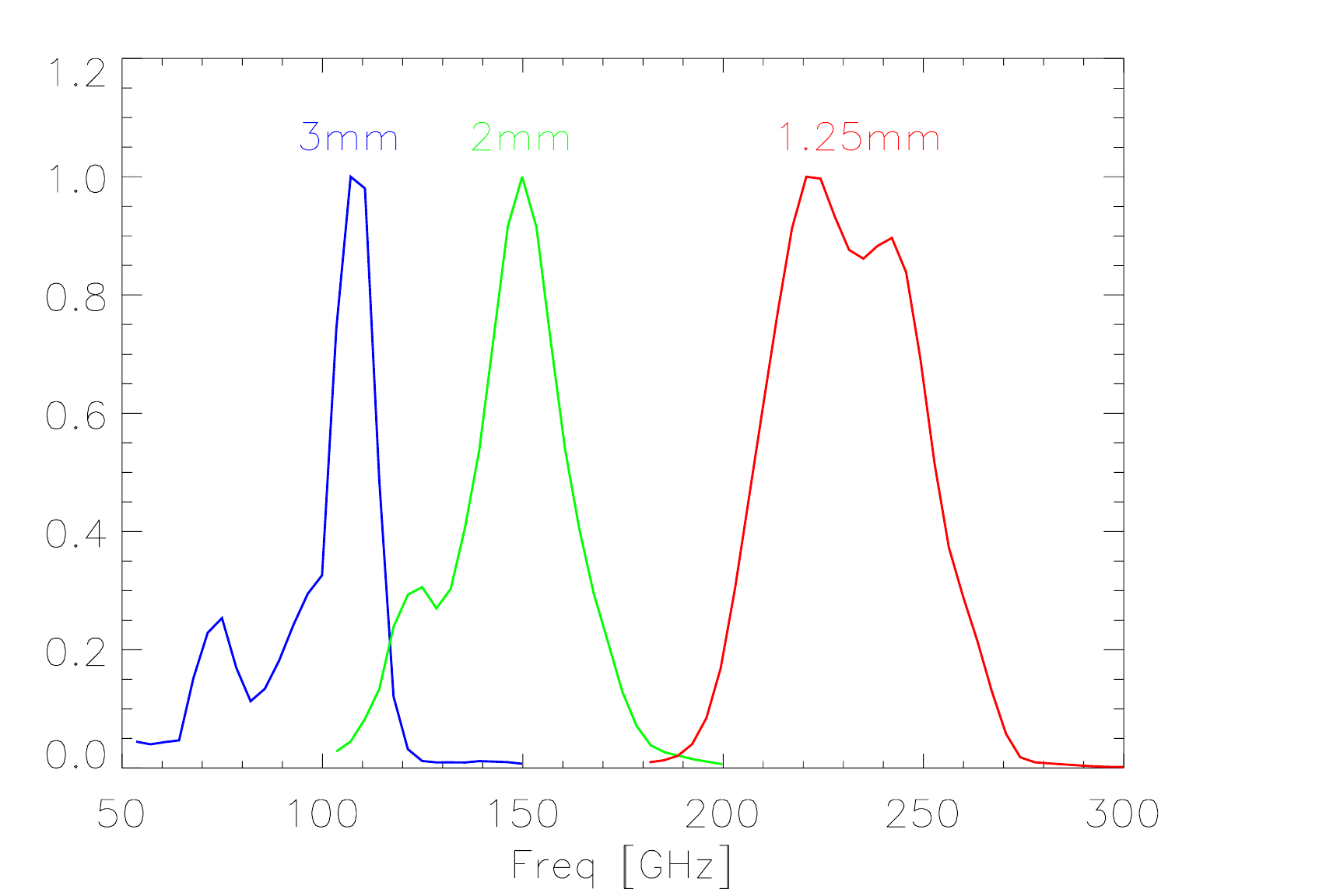}
\includegraphics[trim={3.5cm 14cm 0cm 4cm}, clip=true,scale=0.52]{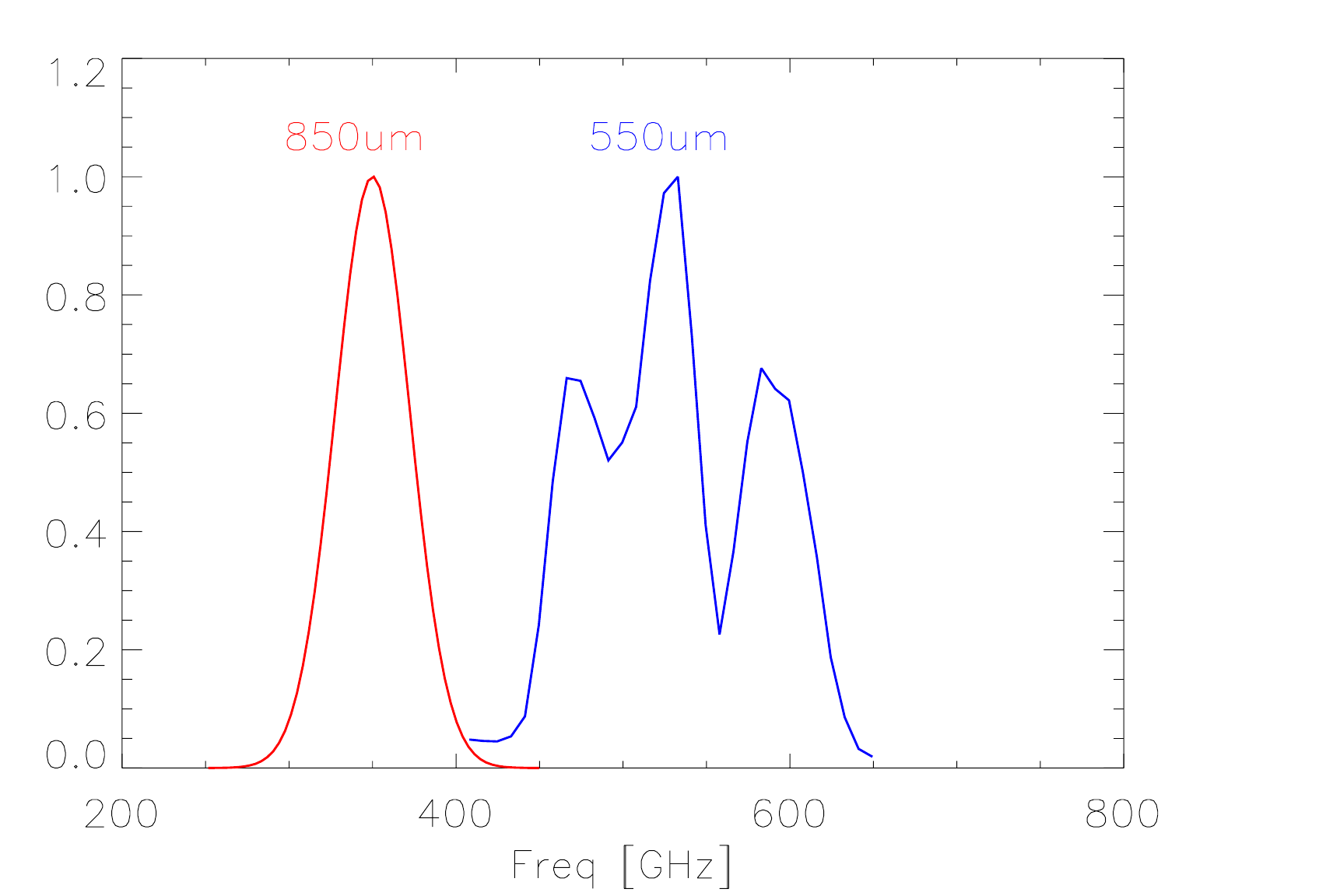}
\end{center}
  \caption{Normalised spectral response for the 5 arrays ( Si/max( Si), where Si is the averaged response of the pixels). The dispersion level between pixels is on the order of a few percent. The dip at 556~GHz is an artifact related to a strong and well-known water vapour absorption line.}
\label{fig6}
\end{figure*}

The measured band transmissions are presented in Fig.~\ref{fig6}. The five spectral bands span the whole range of frequencies from 70~GHz up to 630~GHz, via adjacent bands with a relative FWHM between 15 and 30~\%. The relative bandwidths are mostly limited by the optical filters that were available. The spectral features are well understood and in agreement with 3-D electro-magnetic simulations, when taking into account the full detectors structure. If required by the end application, the spectral curves might be designed to be smoother. For the present study, we have instead optimised the amplitude of the absorption peak.

\subsection{Dynamical response characterisation}\label{subsec4-4}

\begin{figure}[h]
\begin{center}
\includegraphics[width=8.5cm, angle=0]{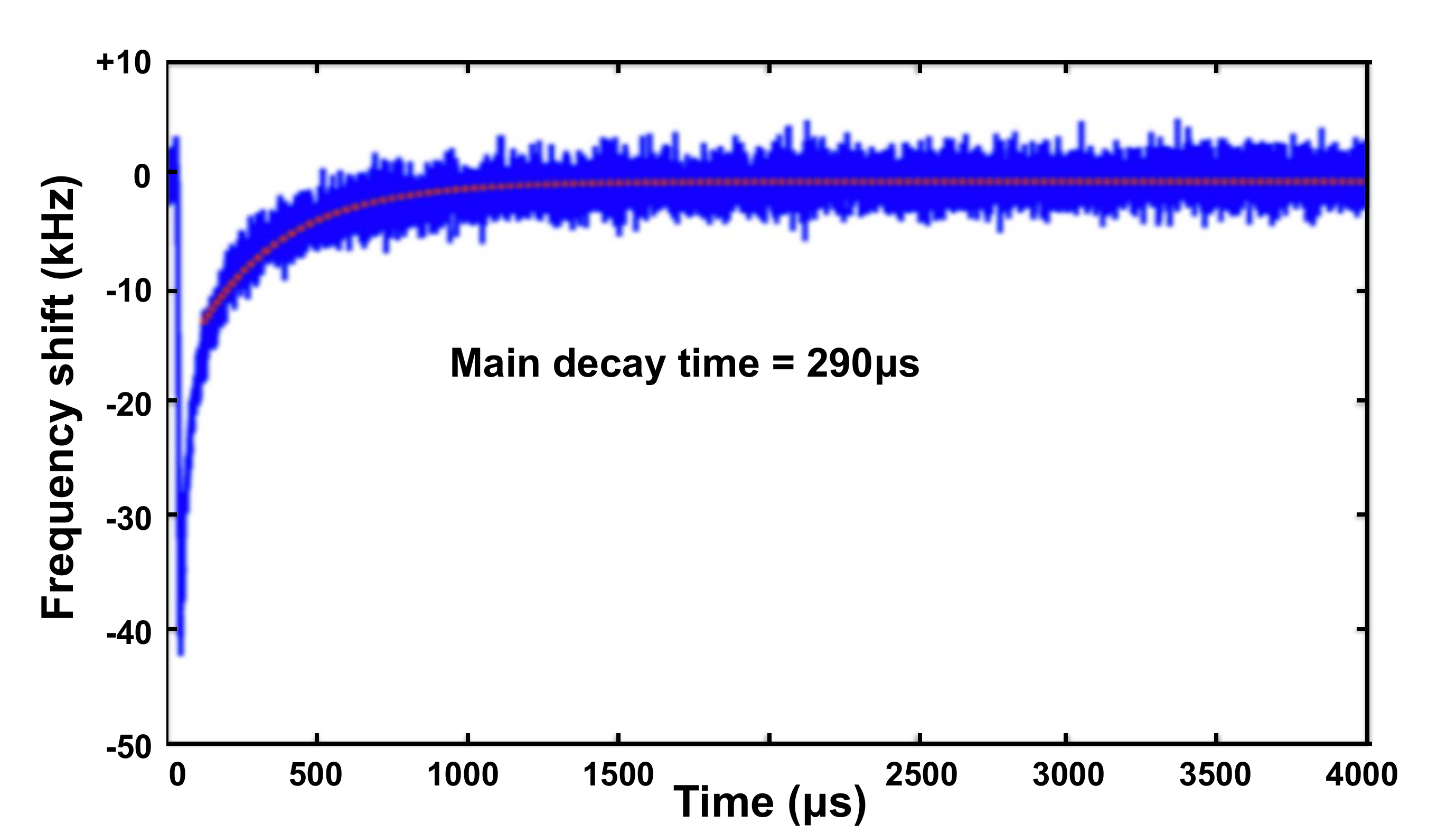}
\end{center}
\caption{One representative cosmic-ray (muon) pulse showing a main decay time of around 0.3~ms. Trace recorded by fast (1~MHZ sampling) homodyne electronics on a typical LEKID of the 3~mm array (Ti-Al film). Array were operated under typical CMB optical background.}
\label{fig7}
\end{figure} 

In order to measure the characteristic response time of LEKID under typical CMB radiation loading, we recorded some fast high-energy events on the 3~mm array. These events are generated by secondary cosmic rays (muons) reaching the ground and crossing the monocrystalline substrate. As a result of this ionising track, a-thermal ballistic phonons are generated in the crystal. A fraction of these phonons is energetic enough to break Cooper pairs in the superconductor film, generating lossy quasi-particles and changing the kinetic inductance of the resonator \citep{swenson2010}. By fitting the decay time of the spikes, we derived the quasi-particles recombination time constant in the Ti-Al film that coincides, for these pair-breaking detectors, with the response time. We present a typical cosmic-ray event in Fig.~\ref{fig7}. The energy deposited by the muon in the substrate is on the order of hundreds of keV. As already shown for example in \citep{2019ApPhL.114c2601K} and \citep{catalano_2}, we know that the impact of cosmic rays in KID detectors is mitigated with respect to thermal detectors by the fact that, when a particle hit occurs, only about 1~\% of the absorbed energy is used to break Cooper pairs and therefore to generate a measurable signal.
Considering the full sample of recorded pulses, and fitting for each decay constant, we measure $\tau_{response}$ = 290 $\pm$ 35~$\mu s$. As expected, this time constant is higher than the one seen on the same kind of pixels operating under higher loading, i.e. higher density of quasi-particles. For example, we have observed, in the past, shorter time constants on the order of tens of microseconds under typical ground-based backgrounds. However, a time constant of around 0.3~ms is still fast enough for a CMB space-borne mission without introducing significant systematic errors \citep{catalano_2, monfardini2016}.

\subsection{Sensitivity characterisation}\label{subsec4-3}

\begin{figure}[h]
\begin{center}
\includegraphics[trim={0cm 0cm 0cm 0cm}, clip=true,scale=0.34]{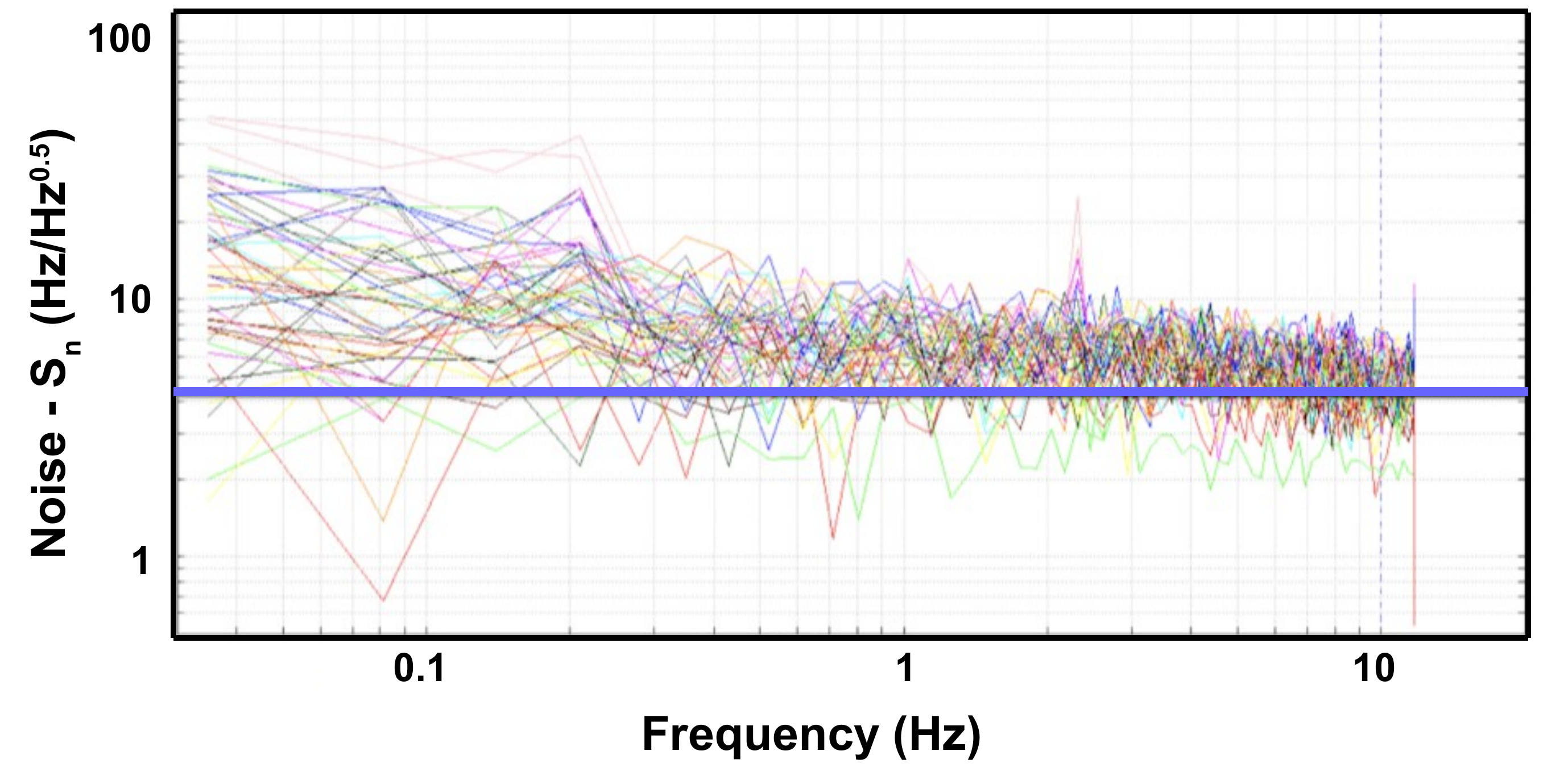}
\end{center}
  \caption{Noise density for all the pixels of the the 2~mm array. Measured with a cold black-body temperature of 6~K. The blue line represents the average noise level after common mode decorrelation.}
\label{fig8}
\end{figure}

\begin{figure*}[h]
\begin{center}
\includegraphics[trim={2cm 14cm 1.5cm 1cm}, clip=true,scale=0.5]{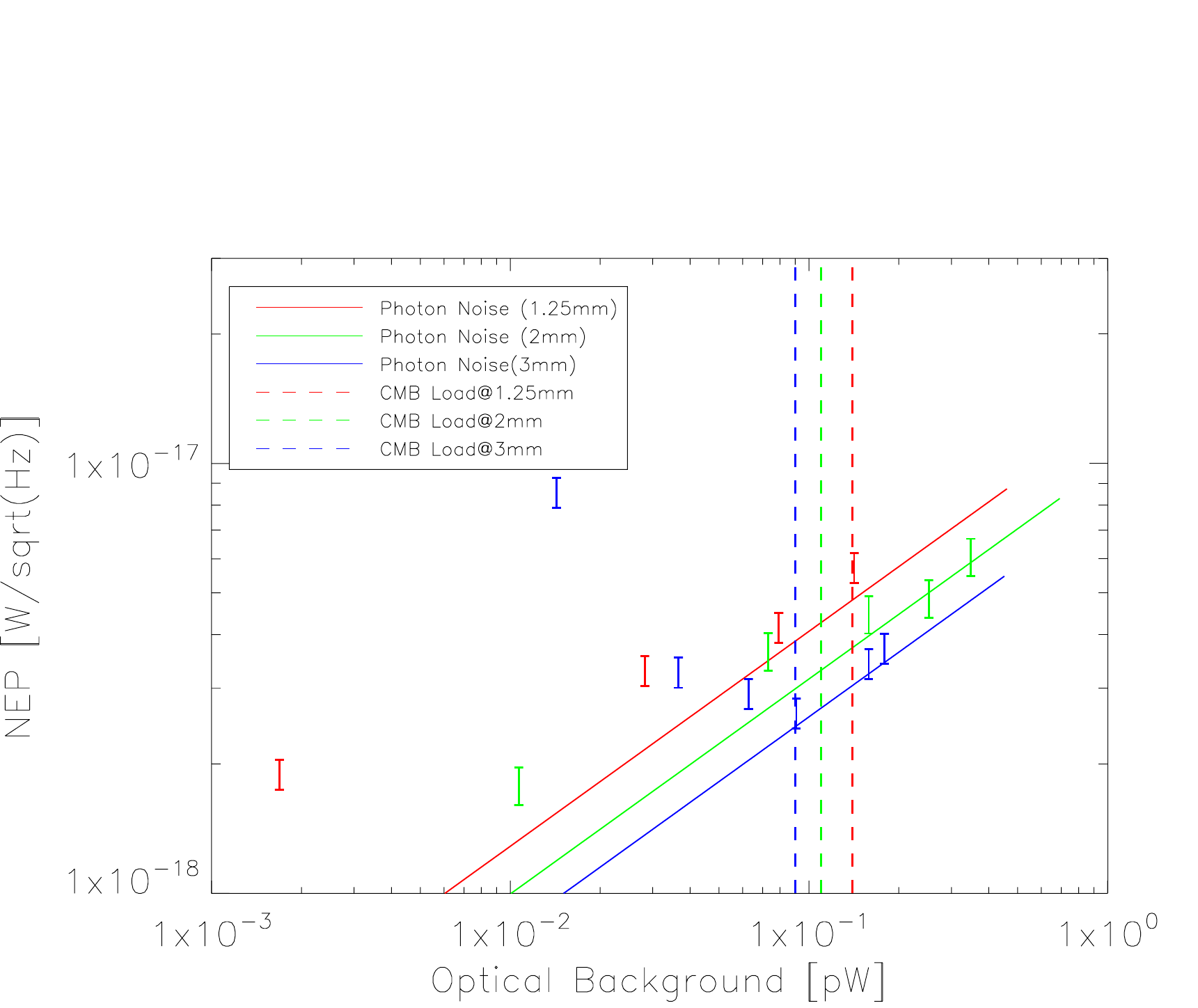}
\includegraphics[trim={1.5cm 14cm 2cm 1cm}, clip=true,scale=0.5]{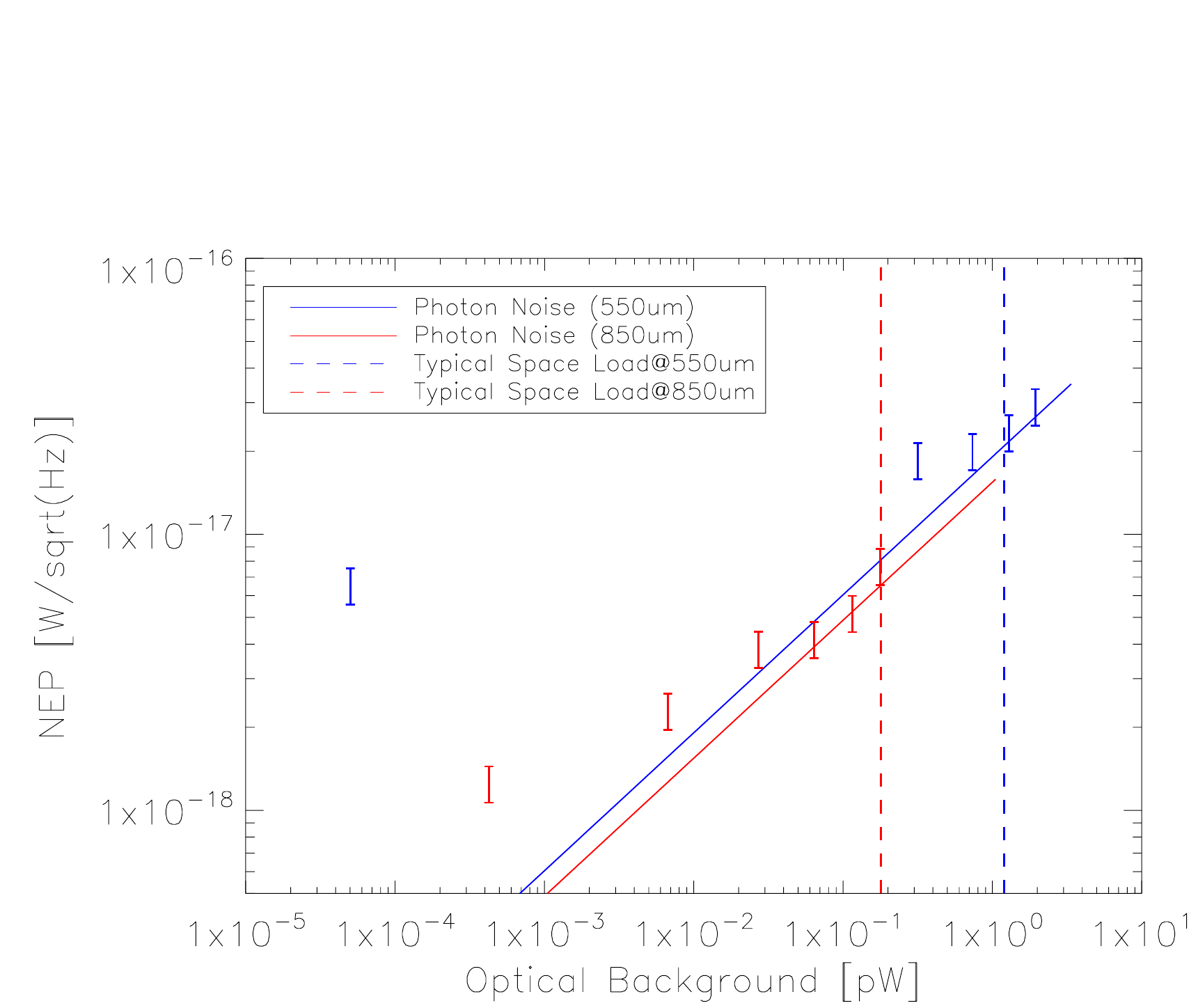}
\end{center}
  \caption{Noise Equivalent Power (NEP) measured for the 1~mm, 2~mm and 3~mm arrays. Each point is derived from responsivity and noise measurements for different temperatures of the cold black-body. The corresponding optical load per pixel for each temperature is derived using the optical simulation presented in Fig.~\ref{fig2}. The vertical dashed lines represent the typical optical background expected from a real instrument operating at second Lagrangian point.}
\label{fig9}
\end{figure*}

The response of each detector (pixel) has been measured by performing, under stable conditions and for several black-body temperatures (T$_{BB}$), frequency sweeps to measure the LEKID transfer function ($\frac{\Delta f}{\Delta T_{BB}}$). The pixel responsivity is then obtained as the ratio between the frequency shift and the corresponding change of the optical background: 

$$
\Re esp (T_{BB}) = \frac{\Delta f}{\Delta W_{opt}}
$$

where $W_{opt}$ is derived from the model presented in Sec.~\ref{sec2}. In parallel, we measured the spectral frequency noise density S$_n$(f), expressed in $Hz / \sqrt{Hz}$, for each T$_{BB}$. In Fig.~\ref{fig8} we show a typical noise (raw) measured for the 2~mm array and with T$_{BB}$=6~K. The correlated electronic noise is removed by subtracting a common mode obtained by averaging the Time-Ordered-Data (TOD) of all detectors. The resulting template is fitted linearly to each TOD. The best-fit is then subtracted from the detector TOD. After decorrelation, the spectral noise density is nicely flat in the band 1-10~Hz. Depending on the array and T$_{BB}$, the frequency noise level sits between 2 and 10~$Hz/\sqrt{Hz}$.

The NEP is then derived as: 

$$
NEP = \frac{S_{n}(f)}{\Re esp}
$$

The Fig.~\ref{fig9} shows the NEP for all the arrays and as a function of the optical background. We compare these results to the corresponding expected $NEP_{phot}$ derived for CMB in a space environment (see also Tab.~\ref{tab2}). 
The results indicate that the NEPs approach the goal (defined in Eq.~\ref{eq1}), in the range of CMB loading, for all the arrays (bands) presented in this paper. For higher optical loads the measured NEP scale correctly. For lower optical loads, on the other hand, the expected NEP scaling with the theoretical photon-noise is not respected. This is likely due to the fact that our optical setup might introduce an additional background when running at fW-per-pixel levels. A deeper investigation of this regime of backgrounds is beyond the scope of the present study.

\section{Conclusion}\label{sec5} 
We have extensively characterised LEKID arrays optimised to work within five bands suitable for CMB space observations.
We have elaborated standard measurement protocols allowing to potentially test a large number of arrays in a reproducible and documented way. This is extremely important in an R\&D phase in which new materials and designs have to be compared in order to identify promising recipes. Adopting this same protocol, we have tested the arrays covering the band 70-630~GHz. 
All the LEKID arrays seem to satisfy the stringent requirements, in terms of sensitivity, of post-Planck satellites dedicated to CMB polarisation studies.

The future generation space instruments devoted to CMB studies will employ thousands pixels. Based on this study, i.e. fully multiplexed 100-pixels-like arrays run under realistic conditions, the scaling to the final number of pixels seems straightforward.  
The next step of the R\&D phase will be therefore to improve the homogeneity within the arrays.

 \begin{table*}
\begin{center}
\begin{tabular}{cccccc}
\hline
\hline
Wavelength & 3~mm  & 2~mm & 1.25~mm & 850~$\mu$m & 550~$\mu$m \\
Frequency & 100~GHz  & 150~GHz & 240~GHz & 360~GHz & 550~GHz \\
\hline \hline
Optical Load [pW/pixel] & 0.09   & 0.11 & 0.14 & 0.18 & 1.2 \\
T$_{BB}$ [K] & 5 & 5 & 8 & 7 & 11 \\
Averaged Optical Resp [kHz/fW] &  2.3  & 0.78 & 1.1 & 0.84 & 0.19 \\
Averaged Noise (1-10~Hz) [$Hz/\sqrt{Hz}$] & 6 & 3 & 6.5 & 5.5 & 4 \\
NEP [$aW/\sqrt{Hz}$] & 2.6$\pm$0.4 & 4.0$\pm$0.4 & 5.5$\pm$0.5 & 5.2$\pm$1.1 & 23$\pm$3.5 \\
NEP$_{Phot}$ [$aW/\sqrt{Hz}$] & 2.45 & 3.4 & 4.8 & 5.0 & 21 \\
\hline \hline
\end{tabular}
\end{center}
\caption{Main results of the paper: optical power, noise and derived NEP for a given optical load corresponding to the typical space environment. The averaged optical response and noise is derived over the best 30~\% of the pixels for the arrays at 3~mm, 2~mm and 1.25~mm. In the case of 850~$\mu$m and 550~$\mu$m arrays we have considered the best 10~\% of the pixels. NEP$_{Phot}$ is calculated from the photon noise at the given optical load.}
\label{tab2}
\end{table*} 

\begin{acknowledgement}
The engineers more involved in the experimental setup development are G. Garde, H. Rodenas, J.-P. Leggeri, M. Grollier, G. Bres, C. Vescovi, J.-P. Scordilis, E. Perbet. We acknowledge the crucial contributions of the whole Cryogenics and Electronics groups at Institut N\'eel and LPSC. The arrays described in this paper have been produced at the PTA Grenoble micro-fabrication facility. This work has been funded by CNES under an R\&T contract; we thank H. Geoffray and P.-G. Tizien. QYT was partially supported by the UChicago-CNRS fund. 
\end{acknowledgement}

\bibliographystyle{aa}
\bibliography{final_version}

\begin{thebibliography}{27}
\expandafter\ifx\csname natexlab\endcsname\relax\def\natexlab#1{#1}\fi

\bibitem[{{Adam} {et~al.}(2018){Adam}, {Adane}, {Ade}, {Andr{\'e}},
  {Andrianasolo}, {Aussel}, {Beelen}, {Beno{\^\i}t}, {Bideaud}, {Billot},
  {Bourrion}, {Bracco}, {Calvo}, {Catalano}, {Coiffard}, {Comis}, {De Petris},
  {D{\'e}sert}, {Doyle}, {Driessen}, {Evans}, {Goupy}, {Kramer}, {Lagache},
  {Leclercq}, {Leggeri}, {Lestrade}, {Mac{\'\i}as-P{\'e}rez}, {Mauskopf},
  {Mayet}, {Maury}, {Monfardini}, {Navarro}, {Pascale}, {Perotto}, {Pisano},
  {Ponthieu}, {Rev{\'e}ret}, {Rigby}, {Ritacco}, {Romero}, {Roussel}, {Ruppin},
  {Schuster}, {Sievers}, {Triqueneaux}, {Tucker}, \& {Zylka}}]{nika2_1}
{Adam}, R., {Adane}, A., {Ade}, P.~A.~R., {et~al.} 2018, \aap, 609, A115

\bibitem[{{Baselmans} {et~al.}(2017){Baselmans}, {Bueno}, {Yates},
  {Yurduseven}, {Llombart}, {Karatsu}, {Baryshev}, {Ferrari}, {Endo}, {Thoen},
  {de Visser}, {Janssen}, {Murugesan}, {Driessen}, {Coiffard},
  {Martin-Pintado}, {Hargrave}, \& {Griffin}}]{2017A&A...601A..89B}
{Baselmans}, J.~J.~A., {Bueno}, J., {Yates}, S.~J.~C., {et~al.} 2017, \aap,
  601, A89

\bibitem[{{Bourrion} {et~al.}(2012){Bourrion}, {Vescovi}, {Bouly}, {Benoit},
  {Calvo}, {Gallin-Martel}, {Macias-Perez}, \& {Monfardini}}]{nikel2012}
{Bourrion}, O., {Vescovi}, C., {Bouly}, J.~L., {et~al.} 2012, Journal of
  Instrumentation, 7, 7014

\bibitem[{{Carlstrom} {et~al.}(2019){Carlstrom}, {Abazajian}, {Addison},
  {Adshead}, {Ahmed}, {Allen}, {Alonso}, {Alvarez}, {Anderson}, {Arnold},
  {Baccigalupi}, {Bailey}, {Barkats}, {Barron}, {Barry}, {Bartlett}, {Basu
  Thakur}, {Battaglia}, {Baxter}, {Bean}, {Bebek}, {Bender}, {Benson},
  {Berger}, {Bhimani}, {Bischoff}, {Bleem}, {Bocquet}, {Boddy}, {Bonato},
  {Bond}, {Borrill}, {Bouchet}, {Brown}, {Bryan}, {Burkhart}, {Buza}, {Byrum},
  {Calabrese}, {Calafut}, {Caldwell}, {Carlstrom}, {Carron}, {Cecil},
  {Challinor}, {Chang}, {Chinone}, {Cho}, {Cooray}, {Crawford}, {Crites},
  {Cukierman}, {Cyr-Racine}, {de Haan}, {de Zotti}, {Delabrouille},
  {Demarteau}, {Devlin}, {Di Valentino}, {Dobbs}, {Duff}, {Duivenvoorden},
  {Dvorkin}, {Edwards}, {Eimer}, {Errard}, {Essinger-Hileman}, {Fabbian},
  {Feng}, {Ferraro}, {Filippini}, {Flauger}, {Flaugher}, {Fraisse}, {Frolov},
  {Galitzki}, {Galli}, {Ganga}, {Gerbino}, {Gilchriese}, {Gluscevic}, {Green},
  {Grin}, {Grohs}, {Gualtieri}, {Guarino}, {Gudmundsson}, {Habib}, {Haller},
  {Halpern}, {Halverson}, {Hanany}, {Harrington}, {Hasegawa}, {Hasselfield},
  {Hazumi}, {Heitmann}, {Henderson}, {Henning}, {Hill}, {Hlo{\v{z}}ek},
  {Holder}, {Holzapfel}, {Hubmayr}, {Huffenberger}, {Huffer}, {Hui}, {Irwin},
  {Johnson}, {Johnstone}, {Jones}, {Karkare}, {Katayama}, {Kerby}, {Kernovsky},
  {Keskitalo}, {Kisner}, {Knox}, {Kosowsky}, {Kovac}, {Kovetz}, {Kuhlmann},
  {Kuo}, {Kurita}, {Kusaka}, {Lahteenmaki}, {Lawrence}, {Lee}, {Lewis}, {Li},
  {Linder}, {Loverde}, {Lowitz}, {Madhavacheril}, {Mantz}, {Matsuda},
  {Mauskopf}, {McMahon}, {Meerburg}, {Melin}, {Meyers}, {Millea}, {Mohr},
  {Moncelsi}, {Mroczkowski}, {Mukherjee}, {Munchmeyer}, {Nagai}, {Nagy},
  {Namikawa}, {Nati}, {Natoli}, {Negrello}, {Newburgh}, {Niemack}, {Nishino},
  {Nordby}, {Novosad}, {O'Connor}, {Obied}, {Padin}, {Pand ey}, {Partridge},
  {Pierpaoli}, {Pogosian}, {Pryke}, {Puglisi}, {Racine}, {Raghunathan},
  {Rahlin}, {Rajagopalan}, {Raveri}, {Reichanadter}, {Reichardt},
  {Remazeilles}, {Rocha}, {Roe}, {Roy}, {Ruhl}, {Salatino}, {Saliwanchik},
  {Schaan}, {Schillaci}, {Schmittfull}, {Scott}, {Sehgal}, {Shandera},
  {Sheehy}, {Sherwin}, {Shirokoff}, {Simon}, {Slosar}, {Somerville}, {Staggs},
  {Stark}, {Stompor}, {Story}, {Stoughton}, {Suzuki}, {Tajima}, {Teply},
  {Thompson}, {Timbie}, {Tomasi}, {Treu}, {Tristram}, {Tucker}, {Umilta}, {van
  Engelen}, {Vieira}, {Vieregg}, {Vogelsberger}, {Wang}, {Watson}, {White},
  {Whitehorn}, {Wollack}, {Wu}, {Xu}, {Yasini}, {Yeck}, {Yoon}, {Young}, \&
  {Zonca}}]{2019BAAS...51g.209C}
{Carlstrom}, J., {Abazajian}, K., {Addison}, G., {et~al.} 2019, in \baas,
  Vol.~51, 209

\bibitem[{{Catalano} {et~al.}(2016){Catalano}, {Benoit}, {Bourrion}, {Calvo},
  {Coiffard}, {D'Addabbo}, {Goupy}, {Le Sueur}, {Mac{\'\i}as-P{\'e}rez}, \&
  {Monfardini}}]{catalano_2}
{Catalano}, A., {Benoit}, A., {Bourrion}, O., {et~al.} 2016, \aap, 592, A26

\bibitem[{{Catalano} {et~al.}(2015){Catalano}, {Goupy}, {le Sueur}, {Benoit},
  {Bourrion}, {Calvo}, {D'addabbo}, {Dumoulin}, {Levy-Bertrand},
  {Mac{\'\i}as-P{\'e}rez}, {Marnieros}, {Ponthieu}, \&
  {Monfardini}}]{catalano_1}
{Catalano}, A., {Goupy}, J., {le Sueur}, H., {et~al.} 2015, \aap, 580, A15

\bibitem[{{de Bernardis} {et~al.}(2018){de Bernardis}, {Ade}, {Baselmans},
  {Battistelli}, {Benoit}, {Bersanelli}, {Bideaud}, {Calvo}, {Casas},
  {Castellano}, {Catalano}, {Charles}, {Colantoni}, {Columbro}, {Coppolecchia},
  {Crook}, {D'Alessandro}, {De Petris}, {Delabrouille}, {Doyle}, {Franceschet},
  {Gomez}, {Goupy}, {Hanany}, {Hills}, {Lamagna}, {Macias-Perez}, {Maffei},
  {Martin}, {Martinez-Gonzalez}, {Masi}, {McCarthy}, {Mennella}, {Monfardini},
  {Noviello}, {Paiella}, {Piacentini}, {Piat}, {Pisano}, {Signorelli}, {Tan},
  {Tartari}, {Trappe}, {Triqueneaux}, {Tucker}, {Vermeulen}, {Young},
  {Zannoni}, {Ach{\'u}carro}, {Allison}, {Artall}, {Ashdown}, {Ballardini},
  {Band ay}, {Banerji}, {Bartlett}, {Bartolo}, {Basak}, {Bonaldi}, {Bonato},
  {Borrill}, {Bouchet}, {Boulanger}, {Brinckmann}, {Bucher}, {Burigana},
  {Buzzelli}, {Cai}, {Carvalho}, {Challinor}, {Chluba}, {Clesse}, {De
  Gasperis}, {De Zotti}, {Di Valentino}, {Diego}, {Errard}, {Feeney},
  {Fernandez-Cobos}, {Finelli}, {Forastieri}, {Galli}, {G{\'e}nova-Santos},
  {Gerbino}, {Gonz{\'a}lez-Nuevo}, {Hagstotz}, {Greenslade}, {Handley},
  {Hern{\'a}ndez-Monteagudo}, {Hervias-Caimapo}, {Hivon}, {Kiiveri}, {Kisner},
  {Kitching}, {Kunz}, {Kurki-Suonio}, {Lasenby}, {Lattanzi}, {Lesgourgues},
  {Lewis}, {Liguori}, {Lindholm}, {Luzzi}, {Martins}, {Matarrese},
  {Melchiorri}, {Melin}, {Molinari}, {Natoli}, {Negrello}, {Notari},
  {Paoletti}, {Patanchon}, {Polastri}, {Polenta}, {Pollo}, {Poulin}, {Quartin},
  {Remazeilles}, {Roman}, {Rubi{\~n}o-Mart{\'\i}n}, {Salvati}, {Tomasi},
  {Tramonte}, {Trombetti}, {V{\"a}liviita}, {Van de Weyjgaert}, {van Tent},
  {Vennin}, {Vielva}, \& {Vittorio}}]{2018JCAP...04..015D}
{de Bernardis}, P., {Ade}, P.~A.~R., {Baselmans}, J.~J.~A., {et~al.} 2018,
  \jcap, 2018, 015

\bibitem[{{Flanigan} {et~al.}(2016){Flanigan}, {McCarrick}, {Jones}, {Johnson},
  {Abitbol}, {Ade}, {Araujo}, {Bradford}, {Cantor}, {Che}, {Day}, {Doyle},
  {Kjellstrand }, {Leduc}, {Limon}, {Luu}, {Mauskopf}, {Miller}, {Mroczkowski},
  {Tucker}, \& {Zmuidzinas}}]{2016ApPhL.108h3504F}
{Flanigan}, D., {McCarrick}, H., {Jones}, G., {et~al.} 2016, Applied Physics
  Letters, 108, 083504

\bibitem[{{Goupy} {et~al.}(2016){Goupy}, {Adane}, {Benoit}, {Bourrion},
  {Calvo}, {Catalano}, {Coiffard}, {Hoarau}, {Leclercq}, {Le Sueur},
  {Macias-Perez}, {Monfardini}, {Peck}, \& {Schuster}}]{dicing2016}
{Goupy}, J., {Adane}, A., {Benoit}, A., {et~al.} 2016, Journal of Low
  Temperature Physics, 184, 661

\bibitem[{{Hu} \& {White}(1997)}]{1997PhRvD..56..596H}
{Hu}, W. \& {White}, M. 1997, \prd, 56, 596

\bibitem[{{Karatsu} {et~al.}(2019){Karatsu}, {Endo}, {Bueno}, {de Visser},
  {Barends}, {Thoen}, {Murugesan}, {Tomita}, \&
  {Baselmans}}]{2019ApPhL.114c2601K}
{Karatsu}, K., {Endo}, A., {Bueno}, J., {et~al.} 2019, Applied Physics Letters,
  114, 032601

\bibitem[{{Leduc} {et~al.}(2010){Leduc}, {Bumble}, {Day}, {Eom}, {Gao},
  {Golwala}, {Mazin}, {McHugh}, {Merrill}, {Moore}, {Noroozian}, {Turner}, \&
  {Zmuidzinas}}]{leduc}
{Leduc}, H.~G., {Bumble}, B., {Day}, P.~K., {et~al.} 2010, Applied Physics
  Letters, 97, 102509

\bibitem[{{Lee} {et~al.}(2019){Lee}, {Ade}, {Akiba}, {Alonso}, {Arnold},
  {Aumont}, {Austermann}, {Baccigalupi}, {Banday}, {Banerji}, {Barreiro},
  {Basak}, {Beall}, {Beckman}, {Bersanelli}, {Borrill}, {Boulanger}, {Brown},
  {Bucher}, {Buzzelli}, {Calabrese}, {Casas}, {Challinor}, {Chan}, {Chinone},
  {Cliche}, {Columbro}, {Cukierman}, {Curtis}, {Danto}, {de Bernardis}, {de
  Haan}, {De Petris}, {Dickinson}, {Dobbs}, {Dotani}, {Duband}, {Ducout},
  {Duff}, {Duivenvoorden}, {Duval}, {Ebisawa}, {Elleflot}, {Enokida},
  {Eriksen}, {Errard}, {Essinger-Hileman}, {Finelli}, {Flauger}, {Franceschet},
  {Fuskeland }, {Ganga}, {Gao}, {Genova-Santos}, {Ghigna}, {Gomez}, {Gradziel},
  {Grain}, {Grupp}, {Gruppuso}, {Gudmundsson}, {Halverson}, {Hargrave},
  {Hasebe}, {Hasegawa}, {Hattori}, {Hazumi}, {Henrot-Versille}, {Herranz},
  {Hill}, {Hilton}, {Hirota}, {Hivon}, {Hlo{\v{z}}ek}, {Hoang}, {Hubmayr},
  {Ichiki}, {Iida}, {Imada}, {Ishimura}, {Ishino}, {Jaehnig}, {Jones}, {Kaga},
  {Kashima}, {Kataoka}, {Katayama}, {Kawasaki}, {Keskitalo}, {Kibayashi},
  {Kikuchi}, {Kimura}, {Kisner}, {Kobayashi}, {Kogiso}, {Kogut}, {Kohri},
  {Komatsu}, {Komatsu}, {Konishi}, {Krachmalnicoff}, {Kuo}, {Kurinsky},
  {Kushino}, {Kuwata-Gonokami}, {Lamagna}, {Lattanzi}, {Lee}, {Linder},
  {Maffei}, {Maino}, {Maki}, {Mangilli}, {Mart{\'\i}nez-Gonzalez}, {Masi},
  {Mathon}, {Matsumura}, {Mennella}, {Migliaccio}, {Minami}, {Mistuda},
  {Molinari}, {Montier}, {Morgante}, {Mot}, {Murata}, {Murphy}, {Nagai},
  {Nagata}, {Nakamura}, {Namikawa}, {Natoli}, {Nerval}, {Nishibori}, {Nishino},
  {Nomura}, {Noviello}, {O'Sullivan}, {Ochi}, {Ogawa}, {Ogawa}, {Ohsaki},
  {Ohta}, {Okada}, {Okada}, {Pagano}, {Paiella}, {Paoletti}, {Patanchon},
  {Piacentini}, {Pisano}, {Polenta}, {Poletti}, {Prouv{\'e}}, {Puglisi},
  {Rambaud}, {Raum}, {Realini}, {Remazeilles}, {Roudil},
  {Rubi{\~n}o-Mart{\i}n}, {Russell}, {Sakurai}, {Sakurai}, {Sandri}, {Savini},
  {Scott}, {Sekimoto}, {Sherwin}, {Shinozaki}, {Shiraishi}, {Shirron},
  {Signorelli}, {Smecher}, {Spizzi}, {Stever}, {Stompor}, {Sugai}, {Sugiyama},
  {Suzuki}, {Suzuki}, {Switzer}, {Takaku}, {Takakura}, {Takakura}, {Takeda},
  {Taylor}, {Taylor}, {Terao}, {Thompson}, {Thorne}, {Tomasi}, {Tomida},
  {Trappe}, {Tristram}, {Tsuji}, {Tsujimoto}, {Tucker}, {Ullom}, {Uozumi},
  {Utsunomiya}, {Van Lanen}, {Vermeulen}, {Vielva}, {Villa}, {Vissers},
  {Vittorio}, {Voisin}, {Walker}, {Watanabe}, {Wehus}, {Weller}, {Westbrook},
  {Winter}, {Wollack}, {Yamamoto}, {Yamasaki}, {Yanagisawa}, {Yoshida},
  {Yumoto}, {Zannoni}, \& {Zonca}}]{2019BAAS...51g.286L}
{Lee}, A., {Ade}, P.~A.~R., {Akiba}, Y., {et~al.} 2019, in \baas, Vol.~51, 286

\bibitem[{{Lee}(2016)}]{2016AAS...22830108L}
{Lee}, A.~T. 2016, in American Astronomical Society Meeting Abstracts, Vol.
  228, American Astronomical Society Meeting Abstracts \#228, 301.08

\bibitem[{{Masi} {et~al.}(2019){Masi}, {de Bernardis}, {Paiella}, {Piacentini},
  {Lamagna}, {Coppolecchia}, {Ade}, {Battistelli}, {Castellano}, {Colantoni},
  {Columbro}, {D'Alessandro}, {De Petris}, {Gordon}, {Magneville}, {Mauskopf},
  {Pettinari}, {Pisano}, {Polenta}, {Presta}, {Tommasi}, {Tucker}, {Vdovin},
  {Volpe}, \& {Yvon}}]{Masi_Olimpo2019}
{Masi}, S., {de Bernardis}, P., {Paiella}, A., {et~al.} 2019, \jcap, 2019, 003

\bibitem[{{Monfardini} {et~al.}(2016){Monfardini}, {Baselmans}, {Benoit},
  {Bideaud}, {Bourrion}, {Catalano}, {Calvo}, {D'Addabbo}, {Doyle}, {Goupy},
  {Le Sueur}, \& {Macias-Perez}}]{monfardini2016}
{Monfardini}, A., {Baselmans}, J., {Benoit}, A., {et~al.} 2016, Society of
  Photo-Optical Instrumentation Engineers (SPIE) Conference Series, Vol. 9914,
  {Lumped element kinetic inductance detectors for space applications}, 99140N

\bibitem[{{Monfardini} {et~al.}(2011){Monfardini}, {Benoit}, {Bideaud},
  {Swenson}, {Cruciani}, {Camus}, {Hoffmann}, {D{\'e}sert}, {Doyle}, {Ade},
  {Mauskopf}, {Tucker}, {Roesch}, {Leclercq}, {Schuster}, {Endo}, {Baryshev},
  {Baselmans}, {Ferrari}, {Yates}, {Bourrion}, {Macias-Perez}, {Vescovi},
  {Calvo}, \& {Giordano}}]{nika1}
{Monfardini}, A., {Benoit}, A., {Bideaud}, A., {et~al.} 2011, \apjs, 194, 24

\bibitem[{{Perotto} {et~al.}(2019){Perotto}, {Ponthieu},
  {Mac{\'\i}as-P{\'e}rez}, {Adam}, {Ade}, {Andr{\'e}}, {Andrianasolo},
  {Aussel}, {Beelen}, {Beno{\^\i}t}, {Berta}, {Bideaud}, {Bourrion}, {Calvo},
  {Catalano}, {Comis}, {De Petris}, {D{\'e}sert}, {Doyle}, {Driessen},
  {Garc{\'\i}a}, {Gomez}, {Goupy}, {John}, {K{\'e}ruzor{\'e}}, {Kramer},
  {Ladjelate}, {Lagache}, {Leclercq}, {Lestrade}, {Maury}, {Mauskopf}, {Mayet},
  {Monfardini}, {Navarro}, {Pe{\~n}alver}, {Pierfederici}, {Pisano},
  {Rev{\'e}ret}, {Ritacco}, {Romero}, {Roussel}, {Ruppin}, {Schuster}, {Shu},
  {Sievers}, {Tucker}, \& {Zylka}}]{nika2_2}
{Perotto}, L., {Ponthieu}, N., {Mac{\'\i}as-P{\'e}rez}, J.~F., {et~al.} 2019,
  arXiv e-prints, arXiv:1910.02038

\bibitem[{{Planck Collaboration} {et~al.}(2016{\natexlab{a}}){Planck
  Collaboration}, {Adam}, {Ade}, {Aghanim}, {Akrami}, {Alves}, {Arg{\"u}eso},
  {Arnaud}, {Arroja}, {Ashdown}, {Aumont}, {Baccigalupi}, {Ballardini}, {Band
  ay}, {Barreiro}, {Bartlett}, {Bartolo}, {Basak}, {Battaglia}, {Battaner},
  {Battye}, {Benabed}, {Beno{\^\i}t}, {Benoit-L{\'e}vy}, {Bernard},
  {Bersanelli}, {Bertincourt}, {Bielewicz}, {Bikmaev}, {Bock}, {B{\"o}hringer},
  {Bonaldi}, {Bonavera}, {Bond}, {Borrill}, {Bouchet}, {Boulanger}, {Bucher},
  {Burenin}, {Burigana}, {Butler}, {Calabrese}, {Cardoso}, {Carvalho},
  {Casaponsa}, {Castex}, {Catalano}, {Challinor}, {Chamballu}, {Chary},
  {Chiang}, {Chluba}, {Chon}, {Christensen}, {Church}, {Clemens}, {Clements},
  {Colombi}, {Colombo}, {Combet}, {Comis}, {Contreras}, {Couchot}, {Coulais},
  {Crill}, {Cruz}, {Curto}, {Cuttaia}, {Danese}, {Davies}, {Davis}, {de
  Bernardis}, {de Rosa}, {de Zotti}, {Delabrouille}, {Delouis}, {D{\'e}sert},
  {Di Valentino}, {Dickinson}, {Diego}, {Dolag}, {Dole}, {Donzelli},
  {Dor{\'e}}, {Douspis}, {Ducout}, {Dunkley}, {Dupac}, {Efstathiou},
  {Eisenhardt}, {Elsner}, {En{\ss}lin}, {Eriksen}, {Falgarone}, {Fantaye},
  {Farhang}, {Feeney}, {Fergusson}, {Fernandez-Cobos}, {Feroz}, {Finelli},
  {Florido}, {Forni}, {Frailis}, {Fraisse}, {Franceschet}, {Franceschi},
  {Frejsel}, {Frolov}, {Galeotta}, {Galli}, {Ganga}, {Gauthier},
  {G{\'e}nova-Santos}, {Gerbino}, {Ghosh}, {Giard}, {Giraud-H{\'e}raud},
  {Giusarma}, {Gjerl{\o}w}, {Gonz{\'a}lez-Nuevo}, {G{\'o}rski}, {Grainge},
  {Gratton}, {Gregorio}, {Gruppuso}, {Gudmundsson}, {Hamann}, {Handley},
  {Hansen}, {Hanson}, {Harrison}, {Heavens}, {Helou}, {Henrot-Versill{\'e}},
  {Hern{\'a}ndez-Monteagudo}, {Herranz}, {Hildebrandt}, {Hivon}, {Hobson},
  {Holmes}, {Hornstrup}, {Hovest}, {Huang}, {Huffenberger}, {Hurier},
  {Ili{\'c}}, {Jaffe}, {Jaffe}, {Jin}, {Jones}, {Juvela}, {Karakci},
  {Keih{\"a}nen}, {Keskitalo}, {Khamitov}, {Kiiveri}, {Kim}, {Kisner},
  {Kneissl}, {Knoche}, {Knox}, {Krachmalnicoff}, {Kunz}, {Kurki-Suonio},
  {Lacasa}, {Lagache}, {L{\"a}hteenm{\"a}ki}, {Lamarre}, {Langer}, {Lasenby},
  {Lattanzi}, {Lawrence}, {Le Jeune}, {Leahy}, {Lellouch}, {Leonardi},
  {Le{\'o}n-Tavares}, {Lesgourgues}, {Levrier}, {Lewis}, {Liguori}, {Lilje},
  {Lilley}, {Linden-V{\o}rnle}, {Lindholm}, {Liu}, {L{\'o}pez-Caniego},
  {Lubin}, {Ma}, {Mac{\'\i}as-P{\'e}rez}, {Maggio}, {Maino}, {Mak},
  {Mandolesi}, {Mangilli}, {Marchini}, {Marcos-Caballero}, {Marinucci},
  {Maris}, {Marshall}, {Martin}, {Martinelli}, {Mart{\'\i}nez-Gonz{\'a}lez},
  {Masi}, {Matarrese}, {Mazzotta}, {McEwen}, {McGehee}, {Mei}, {Meinhold},
  {Melchiorri}, {Melin}, {Mendes}, {Mennella}, {Migliaccio}, {Mikkelsen},
  {Millea}, {Mitra}, {Miville-Desch{\^e}nes}, {Molinari}, {Moneti}, {Montier},
  {Moreno}, {Morgante}, {Mortlock}, {Moss}, {Mottet}, {M{\"u}nchmeyer},
  {Munshi}, {Murphy}, {Narimani}, {Naselsky}, {Nastasi}, {Nati}, {Natoli},
  {Negrello}, {Netterfield}, {N{\o}rgaard-Nielsen}, {Noviello}, {Novikov},
  {Novikov}, {Olamaie}, {Oppermann}, {Orlando}, {Oxborrow}, {Paci}, {Pagano},
  {Pajot}, {Paladini}, {Pandolfi}, {Paoletti}, {Partridge}, {Pasian},
  {Patanchon}, {Pearson}, {Peel}, {Peiris}, {Pelkonen}, {Perdereau}, {Perotto},
  {Perrott}, {Perrotta}, {Pettorino}, {Piacentini}, {Piat}, {Pierpaoli},
  {Pietrobon}, {Plaszczynski}, {Pogosyan}, {Pointecouteau}, {Polenta}, {Popa},
  {Pratt}, {Pr{\'e}zeau}, {Prunet}, {Puget}, {Rachen}, {Racine}, {Reach},
  {Rebolo}, {Reinecke}, {Remazeilles}, {Renault}, {Renzi}, {Ristorcelli},
  {Rocha}, {Roman}, {Romelli}, {Rosset}, {Rossetti}, {Rotti}, {Roudier},
  {Rouill{\'e} d'Orfeuil}, {Rowan-Robinson}, {Rubi{\~n}o-Mart{\'\i}n},
  {Ruiz-Granados}, {Rumsey}, {Rusholme}, {Said}, {Salvatelli}, {Salvati},
  {Sandri}, {Sanghera}, {Santos}, {Saunders}, {Sauv{\'e}}, {Savelainen},
  {Savini}, {Schaefer}, {Schammel}, {Scott}, {Seiffert}, {Serra}, {Shellard},
  {Shimwell}, {Shiraishi}, {Smith}, {Souradeep}, {Spencer}, {Spinelli},
  {Stanford}, {Stern}, {Stolyarov}, {Stompor}, {Strong}, {Sudiwala}, {Sunyaev},
  {Sutter}, {Sutton}, {Suur-Uski}, {Sygnet}, {Tauber}, {Tavagnacco}, {Terenzi},
  {Texier}, {Toffolatti}, {Tomasi}, {Tornikoski}, {Tramonte}, {Tristram},
  {Troja}, {Trombetti}, {Tucci}, {Tuovinen}, {T{\"u}rler}, {Umana},
  {Valenziano}, {Valiviita}, {Van Tent}, {Vassallo}, {Vibert}, {Vidal}, {Viel},
  {Vielva}, {Villa}, {Wade}, {Walter}, {Wand elt}, {Watson}, {Wehus},
  {Welikala}, {Weller}, {White}, {White}, {Wilkinson}, {Yvon}, {Zacchei},
  {Zibin}, \& {Zonca}}]{planck_4}
{Planck Collaboration}, {Adam}, R., {Ade}, P.~A.~R., {et~al.}
  2016{\natexlab{a}}, \aap, 594, A1

\bibitem[{{Planck Collaboration} {et~al.}(2014){Planck Collaboration}, {Ade},
  {Aghanim}, {Armitage-Caplan}, {Arnaud}, {Ashdown}, {Atrio-Barand ela},
  {Aumont}, {Baccigalupi}, {Banday}, {Barreiro}, {Bartlett}, {Battaner},
  {Benabed}, {Beno{\^\i}t}, {Benoit-L{\'e}vy}, {Bernard}, {Bersanelli},
  {Bielewicz}, {Bobin}, {Bock}, {Bonaldi}, {Bond}, {Borrill}, {Bouchet},
  {Bridges}, {Bucher}, {Burigana}, {Butler}, {Calabrese}, {Cappellini},
  {Cardoso}, {Catalano}, {Challinor}, {Chamballu}, {Chary}, {Chen}, {Chiang},
  {Chiang}, {Christensen}, {Church}, {Clements}, {Colombi}, {Colombo},
  {Couchot}, {Coulais}, {Crill}, {Curto}, {Cuttaia}, {Danese}, {Davies},
  {Davis}, {de Bernardis}, {de Rosa}, {de Zotti}, {Delabrouille}, {Delouis},
  {D{\'e}sert}, {Dickinson}, {Diego}, {Dolag}, {Dole}, {Donzelli}, {Dor{\'e}},
  {Douspis}, {Dunkley}, {Dupac}, {Efstathiou}, {Elsner}, {En{\ss}lin},
  {Eriksen}, {Finelli}, {Forni}, {Frailis}, {Fraisse}, {Franceschi}, {Gaier},
  {Galeotta}, {Galli}, {Ganga}, {Giard}, {Giardino}, {Giraud-H{\'e}raud},
  {Gjerl{\o}w}, {Gonz{\'a}lez-Nuevo}, {G{\'o}rski}, {Gratton}, {Gregorio},
  {Gruppuso}, {Gudmundsson}, {Haissinski}, {Hamann}, {Hansen}, {Hanson},
  {Harrison}, {Henrot-Versill{\'e}}, {Hern{\'a}ndez-Monteagudo}, {Herranz},
  {Hildebrand t}, {Hivon}, {Hobson}, {Holmes}, {Hornstrup}, {Hou}, {Hovest},
  {Huffenberger}, {Jaffe}, {Jaffe}, {Jewell}, {Jones}, {Juvela},
  {Keih{\"a}nen}, {Keskitalo}, {Kisner}, {Kneissl}, {Knoche}, {Knox}, {Kunz},
  {Kurki-Suonio}, {Lagache}, {L{\"a}hteenm{\"a}ki}, {Lamarre}, {Lasenby},
  {Lattanzi}, {Laureijs}, {Lawrence}, {Leach}, {Leahy}, {Leonardi},
  {Le{\'o}n-Tavares}, {Lesgourgues}, {Lewis}, {Liguori}, {Lilje},
  {Linden-V{\o}rnle}, {L{\'o}pez-Caniego}, {Lubin}, {Mac{\'\i}as-P{\'e}rez},
  {Maffei}, {Maino}, {Mand olesi}, {Maris}, {Marshall}, {Martin},
  {Mart{\'\i}nez-Gonz{\'a}lez}, {Masi}, {Massardi}, {Matarrese}, {Matthai},
  {Mazzotta}, {Meinhold}, {Melchiorri}, {Melin}, {Mendes}, {Menegoni},
  {Mennella}, {Migliaccio}, {Millea}, {Mitra}, {Miville-Desch{\^e}nes},
  {Moneti}, {Montier}, {Morgante}, {Mortlock}, {Moss}, {Munshi}, {Murphy},
  {Naselsky}, {Nati}, {Natoli}, {Netterfield}, {N{\o}rgaard-Nielsen},
  {Noviello}, {Novikov}, {Novikov}, {O'Dwyer}, {Osborne}, {Oxborrow}, {Paci},
  {Pagano}, {Pajot}, {Paladini}, {Paoletti}, {Partridge}, {Pasian},
  {Patanchon}, {Pearson}, {Pearson}, {Peiris}, {Perdereau}, {Perotto},
  {Perrotta}, {Pettorino}, {Piacentini}, {Piat}, {Pierpaoli}, {Pietrobon},
  {Plaszczynski}, {Platania}, {Pointecouteau}, {Polenta}, {Ponthieu}, {Popa},
  {Poutanen}, {Pratt}, {Pr{\'e}zeau}, {Prunet}, {Puget}, {Rachen}, {Reach},
  {Rebolo}, {Reinecke}, {Remazeilles}, {Renault}, {Ricciardi}, {Riller},
  {Ristorcelli}, {Rocha}, {Rosset}, {Roudier}, {Rowan-Robinson},
  {Rubi{\~n}o-Mart{\'\i}n}, {Rusholme}, {Sandri}, {Santos}, {Savelainen},
  {Savini}, {Scott}, {Seiffert}, {Shellard}, {Spencer}, {Starck}, {Stolyarov},
  {Stompor}, {Sudiwala}, {Sunyaev}, {Sureau}, {Sutton}, {Suur-Uski}, {Sygnet},
  {Tauber}, {Tavagnacco}, {Terenzi}, {Toffolatti}, {Tomasi}, {Tristram},
  {Tucci}, {Tuovinen}, {T{\"u}rler}, {Umana}, {Valenziano}, {Valiviita}, {Van
  Tent}, {Vielva}, {Villa}, {Vittorio}, {Wade}, {Wandelt}, {Wehus}, {White},
  {White}, {Wilkinson}, {Yvon}, {Zacchei}, \& {Zonca}}]{planck_3}
{Planck Collaboration}, {Ade}, P.~A.~R., {Aghanim}, N., {et~al.} 2014, \aap,
  571, A16

\bibitem[{{Planck Collaboration} {et~al.}(2011){Planck Collaboration}, {Ade},
  {Aghanim}, {Arnaud}, {Ashdown}, {Aumont}, {Baccigalupi}, {Baker}, {Balbi},
  {Banday}, {Barreiro}, {Bartlett}, {Battaner}, {Benabed}, {Bennett},
  {Beno{\^\i}t}, {Bernard}, {Bersanelli}, {Bhatia}, {Bock}, {Bonaldi}, {Bond},
  {Borrill}, {Bouchet}, {Bradshaw}, {Bremer}, {Bucher}, {Burigana}, {Butler},
  {Cabella}, {Cantalupo}, {Cappellini}, {Cardoso}, {Carr}, {Casale},
  {Catalano}, {Cay{\'o}n}, {Challinor}, {Chamballu}, {Charra}, {Chary},
  {Chiang}, {Chiang}, {Christensen}, {Clements}, {Colombi}, {Couchot},
  {Coulais}, {Crill}, {Crone}, {Crook}, {Cuttaia}, {Danese}, {D'Arcangelo},
  {Davies}, {Davis}, {de Bernardis}, {de Bruin}, {de Gasperis}, {de Rosa}, {de
  Zotti}, {Delabrouille}, {Delouis}, {D{\'e}sert}, {Dick}, {Dickinson},
  {Dolag}, {Dole}, {Donzelli}, {Dor{\'e}}, {D{\"o}rl}, {Douspis}, {Dupac},
  {Efstathiou}, {En{\ss}lin}, {Eriksen}, {Finelli}, {Foley}, {Forni},
  {Fosalba}, {Frailis}, {Franceschi}, {Freschi}, {Gaier}, {Galeotta},
  {Gallegos}, {Gandolfo}, {Ganga}, {Giard}, {Giardino}, {Gienger},
  {Giraud-H{\'e}raud}, {Gonz{\'a}lez}, {Gonz{\'a}lez-Nuevo}, {G{\'o}rski},
  {Gratton}, {Gregorio}, {Gruppuso}, {Guyot}, {Haissinski}, {Hansen},
  {Harrison}, {Helou}, {Henrot-Versill{\'e}}, {Hern{\'a}ndez-Monteagudo},
  {Herranz}, {Hildebrandt}, {Hivon}, {Hobson}, {Holmes}, {Hornstrup}, {Hovest},
  {Hoyland}, {Huffenberger}, {Jaffe}, {Jagemann}, {Jones}, {Juillet}, {Juvela},
  {Kangaslahti}, {Keih{\"a}nen}, {Keskitalo}, {Kisner}, {Kneissl}, {Knox},
  {Krassenburg}, {Kurki-Suonio}, {Lagache}, {L{\"a}hteenm{\"a}ki}, {Lamarre},
  {Lange}, {Lasenby}, {Laureijs}, {Lawrence}, {Leach}, {Leahy}, {Leonardi},
  {Leroy}, {Lilje}, {Linden-V{\o}rnle}, {L{\'o}pez-Caniego}, {Lowe}, {Lubin},
  {Mac{\'\i}as-P{\'e}rez}, {Maciaszek}, {MacTavish}, {Maffei}, {Maino},
  {Mandolesi}, {Mann}, {Maris}, {Mart{\'\i}nez-Gonz{\'a}lez}, {Masi},
  {Massardi}, {Matarrese}, {Matthai}, {Mazzotta}, {McDonald}, {McGehee},
  {Meinhold}, {Melchiorri}, {Melin}, {Mendes}, {Mennella}, {Mevi},
  {Miniscalco}, {Mitra}, {Miville-Desch{\^e}nes}, {Moneti}, {Montier},
  {Morgante}, {Morisset}, {Mortlock}, {Munshi}, {Murphy}, {Naselsky}, {Natoli},
  {Netterfield}, {N{\o}rgaard-Nielsen}, {Noviello}, {Novikov}, {Novikov},
  {O'Dwyer}, {Ortiz}, {Osborne}, {Osuna}, {Oxborrow}, {Pajot}, {Paladini},
  {Partridge}, {Pasian}, {Passvogel}, {Patanchon}, {Pearson}, {Pearson},
  {Perdereau}, {Perotto}, {Perrotta}, {Piacentini}, {Piat}, {Pierpaoli},
  {Plaszczynski}, {Platania}, {Pointecouteau}, {Polenta}, {Ponthieu}, {Popa},
  {Poutanen}, {Pr{\'e}zeau}, {Prunet}, {Puget}, {Rachen}, {Reach}, {Rebolo},
  {Reinecke}, {Reix}, {Renault}, {Ricciardi}, {Riller}, {Ristorcelli}, {Rocha},
  {Rosset}, {Rowan-Robinson}, {Rubi{\~n}o-Mart{\'\i}n}, {Rusholme}, {Salerno},
  {Sandri}, {Santos}, {Savini}, {Schaefer}, {Scott}, {Seiffert}, {Shellard},
  {Simonetto}, {Smoot}, {Sozzi}, {Starck}, {Sternberg}, {Stivoli}, {Stolyarov},
  {Stompor}, {Stringhetti}, {Sudiwala}, {Sunyaev}, {Sygnet}, {Tapiador},
  {Tauber}, {Tavagnacco}, {Taylor}, {Terenzi}, {Texier}, {Toffolatti},
  {Tomasi}, {Torre}, {Tristram}, {Tuovinen}, {T{\"u}rler}, {Tuttlebee},
  {Umana}, {Valenziano}, {Valiviita}, {Varis}, {Vibert}, {Vielva}, {Villa},
  {Vittorio}, {Wade}, {Wandelt}, {Watson}, {White}, {White}, {Wilkinson},
  {Yvon}, {Zacchei}, \& {Zonca}}]{planck_1}
{Planck Collaboration}, {Ade}, P.~A.~R., {Aghanim}, N., {et~al.} 2011, \aap,
  536, A1

\bibitem[{{Planck Collaboration} {et~al.}(2016{\natexlab{b}}){Planck
  Collaboration}, {Ade}, {Aghanim}, {Arnaud}, {Ashdown}, {Aumont},
  {Baccigalupi}, {Banday}, {Barreiro}, {Bartlett}, {Bartolo}, {Battaner},
  {Battye}, {Benabed}, {Beno{\^\i}t}, {Benoit-L{\'e}vy}, {Bernard},
  {Bersanelli}, {Bielewicz}, {Bock}, {Bonaldi}, {Bonavera}, {Bond}, {Borrill},
  {Bouchet}, {Boulanger}, {Bucher}, {Burigana}, {Butler}, {Calabrese},
  {Cardoso}, {Catalano}, {Challinor}, {Chamballu}, {Chary}, {Chiang}, {Chluba},
  {Christensen}, {Church}, {Clements}, {Colombi}, {Colombo}, {Combet},
  {Coulais}, {Crill}, {Curto}, {Cuttaia}, {Danese}, {Davies}, {Davis}, {de
  Bernardis}, {de Rosa}, {de Zotti}, {Delabrouille}, {D{\'e}sert}, {Di
  Valentino}, {Dickinson}, {Diego}, {Dolag}, {Dole}, {Donzelli}, {Dor{\'e}},
  {Douspis}, {Ducout}, {Dunkley}, {Dupac}, {Efstathiou}, {Elsner},
  {En{\ss}lin}, {Eriksen}, {Farhang}, {Fergusson}, {Finelli}, {Forni},
  {Frailis}, {Fraisse}, {Franceschi}, {Frejsel}, {Galeotta}, {Galli}, {Ganga},
  {Gauthier}, {Gerbino}, {Ghosh}, {Giard}, {Giraud-H{\'e}raud}, {Giusarma},
  {Gjerl{\o}w}, {Gonz{\'a}lez-Nuevo}, {G{\'o}rski}, {Gratton}, {Gregorio},
  {Gruppuso}, {Gudmundsson}, {Hamann}, {Hansen}, {Hanson}, {Harrison}, {Helou},
  {Henrot-Versill{\'e}}, {Hern{\'a}ndez-Monteagudo}, {Herranz}, {Hildebrand t},
  {Hivon}, {Hobson}, {Holmes}, {Hornstrup}, {Hovest}, {Huang}, {Huffenberger},
  {Hurier}, {Jaffe}, {Jaffe}, {Jones}, {Juvela}, {Keih{\"a}nen}, {Keskitalo},
  {Kisner}, {Kneissl}, {Knoche}, {Knox}, {Kunz}, {Kurki-Suonio}, {Lagache},
  {L{\"a}hteenm{\"a}ki}, {Lamarre}, {Lasenby}, {Lattanzi}, {Lawrence}, {Leahy},
  {Leonardi}, {Lesgourgues}, {Levrier}, {Lewis}, {Liguori}, {Lilje},
  {Linden-V{\o}rnle}, {L{\'o}pez-Caniego}, {Lubin}, {Mac{\'\i}as-P{\'e}rez},
  {Maggio}, {Maino}, {Mandolesi}, {Mangilli}, {Marchini}, {Maris}, {Martin},
  {Martinelli}, {Mart{\'\i}nez-Gonz{\'a}lez}, {Masi}, {Matarrese}, {McGehee},
  {Meinhold}, {Melchiorri}, {Melin}, {Mendes}, {Mennella}, {Migliaccio},
  {Millea}, {Mitra}, {Miville-Desch{\^e}nes}, {Moneti}, {Montier}, {Morgante},
  {Mortlock}, {Moss}, {Munshi}, {Murphy}, {Naselsky}, {Nati}, {Natoli},
  {Netterfield}, {N{\o}rgaard-Nielsen}, {Noviello}, {Novikov}, {Novikov},
  {Oxborrow}, {Paci}, {Pagano}, {Pajot}, {Paladini}, {Paoletti}, {Partridge},
  {Pasian}, {Patanchon}, {Pearson}, {Perdereau}, {Perotto}, {Perrotta},
  {Pettorino}, {Piacentini}, {Piat}, {Pierpaoli}, {Pietrobon}, {Plaszczynski},
  {Pointecouteau}, {Polenta}, {Popa}, {Pratt}, {Pr{\'e}zeau}, {Prunet},
  {Puget}, {Rachen}, {Reach}, {Rebolo}, {Reinecke}, {Remazeilles}, {Renault},
  {Renzi}, {Ristorcelli}, {Rocha}, {Rosset}, {Rossetti}, {Roudier},
  {Rouill{\'e} d'Orfeuil}, {Rowan-Robinson}, {Rubi{\~n}o-Mart{\'\i}n},
  {Rusholme}, {Said}, {Salvatelli}, {Salvati}, {Sandri}, {Santos},
  {Savelainen}, {Savini}, {Scott}, {Seiffert}, {Serra}, {Shellard}, {Spencer},
  {Spinelli}, {Stolyarov}, {Stompor}, {Sudiwala}, {Sunyaev}, {Sutton},
  {Suur-Uski}, {Sygnet}, {Tauber}, {Terenzi}, {Toffolatti}, {Tomasi},
  {Tristram}, {Trombetti}, {Tucci}, {Tuovinen}, {T{\"u}rler}, {Umana},
  {Valenziano}, {Valiviita}, {Van Tent}, {Vielva}, {Villa}, {Wade}, {Wandelt},
  {Wehus}, {White}, {White}, {Wilkinson}, {Yvon}, {Zacchei}, \&
  {Zonca}}]{planck_2}
{Planck Collaboration}, {Ade}, P.~A.~R., {Aghanim}, N., {et~al.}
  2016{\natexlab{b}}, \aap, 594, A13

\bibitem[{{Planck Collaboration} {et~al.}(2018){Planck Collaboration},
  {Akrami}, {Arroja}, {Ashdown}, {Aumont}, {Baccigalupi}, {Ballardini},
  {Banday}, {Barreiro}, {Bartolo}, {Basak}, {Benabed}, {Bernard}, {Bersanelli},
  {Bielewicz}, {Bock}, {Bond}, {Borrill}, {Bouchet}, {Boulanger}, {Bucher},
  {Burigana}, {Butler}, {Calabrese}, {Cardoso}, {Carron}, {Challinor},
  {Chiang}, {Colombo}, {Combet}, {Contreras}, {Crill}, {Cuttaia}, {de
  Bernardis}, {de Zotti}, {Delabrouille}, {Delouis}, {Di Valentino}, {Diego},
  {Donzelli}, {Dor{\'e}}, {Douspis}, {Ducout}, {Dupac}, {Dusini}, {Efstathiou},
  {Elsner}, {En{\ss}lin}, {Eriksen}, {Fantaye}, {Fergusson}, {Fernandez-Cobos},
  {Finelli}, {Forastieri}, {Frailis}, {Franceschi}, {Frolov}, {Galeotta},
  {Galli}, {Ganga}, {Gauthier}, {G{\'e}nova-Santos}, {Gerbino}, {Ghosh},
  {Gonz{\'a}lez-Nuevo}, {G{\'o}rski}, {Gratton}, {Gruppuso}, {Gudmundsson},
  {Hamann}, {Handley}, {Hansen}, {Herranz}, {Hivon}, {Hooper}, {Huang},
  {Jaffe}, {Jones}, {Keih{\"a}nen}, {Keskitalo}, {Kiiveri}, {Kim}, {Kisner},
  {Krachmalnicoff}, {Kunz}, {Kurki-Suonio}, {Lagache}, {Lamarre}, {Lasenby},
  {Lattanzi}, {Lawrence}, {Le Jeune}, {Lesgourgues}, {Levrier}, {Lewis},
  {Liguori}, {Lilje}, {Lindholm}, {Lpez-Caniego}, {Lubin}, {Ma},
  {Mac{\'\i}as-P{\'e}rez}, {Maggio}, {Maino}, {Mandolesi}, {Mangilli},
  {Marcos-Caballero}, {Maris}, {Martin}, {Mart{\'\i}nez-Gonz{\'a}lez},
  {Matarrese}, {Mauri}, {McEwen}, {Meerburg}, {Meinhold}, {Melchiorri},
  {Mennella}, {Migliaccio}, {Mitra}, {Miville-Desch{\^e}nes}, {Molinari},
  {Moneti}, {Montier}, {Morgante}, {Moss}, {M{\"u}nchmeyer}, {Natoli},
  {N{\o}rgaard-Nielsen}, {Pagano}, {Paoletti}, {Partridge}, {Patanchon},
  {Peiris}, {Perrotta}, {Pettorino}, {Piacentini}, {Polastri}, {Polenta},
  {Puget}, {Rachen}, {Reinecke}, {Remazeilles}, {Renzi}, {Rocha}, {Rosset},
  {Roudier}, {Rubi{\~n}o-Mart{\'\i}n}, {Ruiz-Granados}, {Salvati}, {Sandri},
  {Savelainen}, {Scott}, {Shellard}, {Shiraishi}, {Sirignano}, {Sirri},
  {Spencer}, {Sunyaev}, {Suur-Uski}, {Tauber}, {Tavagnacco}, {Tenti},
  {Toffolatti}, {Tomasi}, {Trombetti}, {Valiviita}, {Van Tent}, {Vielva},
  {Villa}, {Vittorio}, {Wandelt}, {Wehus}, {White}, {Zacchei}, {Zibin}, \&
  {Zonca}}]{planck_5}
{Planck Collaboration}, {Akrami}, Y., {Arroja}, F., {et~al.} 2018, arXiv
  e-prints, arXiv:1807.06211

\bibitem[{{Planck Collaboration} \& {Lawrence}(2011)}]{planck_6}
{Planck Collaboration} \& {Lawrence}, C.~R. 2011, in American Astronomical
  Society Meeting Abstracts, Vol. 217, American Astronomical Society Meeting
  Abstracts \#217, 243.04

\bibitem[{{Roesch} {et~al.}(2012){Roesch}, {Benoit}, {Bideaud}, {Boudou},
  {Calvo}, {Cruciani}, {Doyle}, {Leduc}, {Monfardini}, {Swenson}, {Leclercq},
  {Mauskopf}, \& {Schuster}}]{Roesch}
{Roesch}, M., {Benoit}, A., {Bideaud}, A., {et~al.} 2012, arXiv e-prints,
  arXiv:1212.4585

\bibitem[{{Swenson} {et~al.}(2010){Swenson}, {Cruciani}, {Benoit}, {Roesch},
  {Yung}, {Bideaud}, \& {Monfardini}}]{swenson2010}
{Swenson}, L.~J., {Cruciani}, A., {Benoit}, A., {et~al.} 2010, Applied Physics
  Letters, 96, 263511

\bibitem[{{Tsan} \& {Simons Observatory
  Collaboration}(2020)}]{2020AAS...23523504T}
{Tsan}, T. \& {Simons Observatory Collaboration}. 2020, in American
  Astronomical Society Meeting Abstracts, American Astronomical Society Meeting
  Abstracts, 235.04

\end{thebibliography}

\end{document}